%% file: draft.tex
\documentclass[fleqn,usenatbib]{mnras}
\usepackage{newtxtext,newtxmath}
\usepackage[T1]{fontenc}
\usepackage{ae,aecompl}

\usepackage{lastpage, graphicx, amssymb, aas_macros, bm, nicefrac}
\DeclareMathOperator\arcsinh{arcsinh}
\usepackage{bigints}
\usepackage{tabularx}
\usepackage{colortbl}
\usepackage{dcolumn}
\usepackage{xcolor}
\usepackage{ulem}
\hypersetup{breaklinks, colorlinks=true, urlcolor=blue, citecolor=blue, linkcolor=black}
\newcommand{\tuniv}{t_0}
\newcommand{\om}{\Omega_{\rm m}}
\newcommand{\orad}{\Omega_{\rm r}}
\newcommand{\ol}{\Omega_{\Lambda}}

\newcommand{\tlb}{t_{\rm lb}}
\newcommand{\ttilde}{\widetilde{t}}
\newcommand{\planck}{Planck}
\newcommand{\ssim}{{\sim}}

\definecolor{Gray}{gray}{0.9}
\definecolor{update}{HTML}{800000}

\defcitealias{riess2021}{R21}
\defcitealias{poulin2018}{P18}
\defcitealias{chaboyer2017}{C17}

\input{extra_preamble.tex}

\renewcommand{\la}{\lesssim}

\title[Uncertainty in the redshift--time relation]
{Uncertain Times: The Redshift--Time Relation from Cosmology and Stars}
\author[M. Boylan-Kolchin \& D. R. Weisz]
{Michael Boylan-Kolchin$^1$\thanks{$\!$mbk@astro.as.utexas.edu}
  and Daniel R. Weisz$^{2}$\\
\noindent $\!\!^1$Department of Astronomy, The University of Texas at Austin,
2515 Speedway, Stop C1400, Austin, TX 78712-1205, USA\\
\noindent $\!\!^2$Department of Astronomy, University of California Berkeley, Berkeley, CA 94720, USA\\
}

\date{Accepted 2021 May 21. Received 2021 May 19; in original form 2021 March 31}

\pubyear{2021}

\begin{document}
\label{firstpage}
\pagerange{\pageref{firstpage}--\pageref{LastPage}}
\maketitle

\begin{abstract}
\textit{Planck} data provide precise constraints on cosmological parameters when assuming the base \lcdm~model, including a 0.17\% measurement of the age of the Universe, $\tuniv=13.797 \pm 0.023\,\gyr$. 
However, the persistence of the ``Hubble tension" calls the base \lcdm~model's completeness into question and has spurred interest in models such as Early Dark Energy (EDE) that modify the assumed expansion history of the Universe. 
We investigate the effect of EDE on the redshift-time relation $z \leftrightarrow t$ and find that it differs from the base \lcdm\ model by at least ${\approx} 4\%$ at all $t$ and $z$. As long as EDE remains observationally viable, any inferred $t \leftarrow z$ or $z \leftarrow t$ 
quoted to a higher level of precision do not reflect the current status of our understanding of cosmology. 
This uncertainty has important astrophysical implications: the reionization epoch --- $10>z>6$ --- corresponds to disjoint lookback time periods in the base \lcdm~and EDE models,  
and the EDE value of $\tuniv=13.25 \pm 0.17~\gyr$ is in tension with published ages of some stars, star clusters, and ultra-faint dwarf galaxies.
However, most published stellar ages do not include an uncertainty in accuracy 
(due to, e.g., uncertain distances and stellar physics) 
that is estimated to be $\sim7-10\%$, potentially reconciling stellar ages with $t_{0,\rm EDE}$. 
We discuss how the big data era for stars is providing extremely precise ages ($<1\%$) and how improved distances and treatment of stellar physics such as convection could result in ages accurate to $4-5\%$, comparable to the current accuracy of $t \leftrightarrow z$. Such precise and accurate stellar ages can provide detailed insight into the high-redshift Universe independent of a cosmological model. 
\end{abstract}

\begin{keywords}
  cosmological parameters -- distance scale -- cosmic background radiation -- stars: fundamental parameters
\end{keywords}

\section{Introduction}
The basis of observational cosmology is that the finite speed of light means that observations of more distant objects reveal properties of these objects at earlier times in the evolution of the Universe. And yet, neither distance ($d$) nor time ($t$) is a cosmological observable: it is the redshift ($z$) of a galaxy that is  measured, and relating $z$ to $t$ or $d$ requires a cosmological model.
In the context of the base dark energy ($\Lambda$) plus cold dark matter (CDM) model (\lcdm), where it is assumed that dark energy is a cosmological constant and the Universe is spatially flat, ages and cosmological distance measures (or conformal times) at a given redshift depend only on $H_0$ and $\om$.
The precision of age or distance determinations is therefore fundamentally linked to the precision of $H_0$ and $\om$ measurements.

Fortunately --- as untold papers, talks, and press releases remind us --- we live in the age of precision cosmology. The baseline \lcdm\ fit to \textit{Planck} observations of the Cosmic Microwave Background (CMB) measures $H_0=67.32 \pm 0.54 \,\kms\,\mpc^{-1}$ and $\om=0.3153 \pm 0.0073$ \citep{planck2020}; these values, in turn, establish the redshift-age relation precisely. A specific and important example is the age of the Universe, $\tuniv$, which is calculated to be $13.797 \pm 0.023$~Gyr on the basis of \lcdm-based fits to \textit{Planck} data. Upon seeing this number and associated error bar, indicating that the age of the Universe is known to better than 0.2\% precision, a reader would be excused for
thinking cosmological ages are \textit{very} well-known and unworthy
of further scrutiny. 

And yet, there is certainly reason for skepticism about this conclusion. The CMB does not provide a direct measurement of $H_0$ (or $H$ at any redshift): $H_0$ is a so-called derived quantity in CMB analyses that is inferred by measuring other quantities directly. 
\citet[hereafter \citetalias{riess2021}]{riess2021} have measured $H_0=73.2 \pm 1.3\,\kms\,\mpc^{-1}$ using the luminosity distances to type Ia supernovae, calibrated by Cepheid variable stars  with a period-luminosity relationship that is anchored in geometric distances to nearby ($D < 10\,\mpc$) stars and galaxies. This local value of $H_0$ is formally $4.2\,\sigma$ discrepant with the \textit{Planck} result. 
Fixing all other cosmological parameters --- $\om$ is the parameter of primary importance, as discussed in \S~\ref{sec:cosmology} --- 
the age of the Universe scales inversely with
the Hubble constant, $\tuniv\,H_0={\rm constant}$, indicating a possible 10\%
systematic uncertainty in $\tuniv$ and in the age-redshift relation overall.\footnote{The local determination of $H_0$ is not sensitive to $\om$, but changing $H_0$ to the \citetalias{riess2021} value while fixing $\om$ to the base \lcdm~\textit{Planck} value would result in an atrocious fit to CMB data. See \S~\ref{subsec:correlations} for further discussion.}

Although Cepheid-based determinations are perhaps the most well-known way to measure $H_0$ in the local Universe, several different techniques are now being employed (for a recent compilation, see \citealt{di-valentino2021}). 
It is important that the multiple ``late-time'' probes of the expansion rate are available, as systematic errors are different (or even independent) for the different measurements. Even just within the past year, measurements of $H_0$ that rely on the tip of the red giant branch \citep{freedman2020}, Mira variables \citep{huang2020}, surface brightness fluctuations \citep{khetan2021, blakeslee2021}, the Tully-Fisher relation \citep{kourkchi2020, schombert2020}, masers \citep{pesce2020}, and gravitational lenses \citep{birrer2020} have been published, with gravitational wave measurements looming on the horizon as a potentially powerful way to measure the expansion rate in the nearby Universe \citep{holz2005, abbott2017a}. The global picture based on late-time measurements is somewhat murky and is evolving quickly, but it is clear that these late-time determinations are all \textit{no lower} than early-time measurements, and that they are generally higher. For example, by combining over twenty of these late-time measurements, \citet{di-valentino2021} finds $H_0=72.7 \pm 1.1\,{\rm km\,s^{-1}\,Mpc^{-1}}$. 

At the same time, CMB observations from ground-based observatories are obtaining $H_0$ values that agree with the results of the \textit{Planck} satellite. The Atacama Cosmology Telescope (ACT) collaboration finds $H_0=67.9 \pm 1.5 \,{\rm km\,s^{-1}\,Mpc^{-1}}$ from ACT data alone \citep{aiola2020}, while the SPT-3G collaboration recently reported $H_0=68.8 \pm 1.5 \,{\rm km\,s^{-1}\,Mpc^{-1}}$ based solely on SPT-3G $E$-mode polarization autocorrelation and temperature-$E$ cross-correlation functions \citep{dutcher2021}. It is also possible to measure the sound horizon $r_{\rm d}$ at the end of the baryon drag epoch, $z_{\rm d} \!\approx\! 1060$, via the baryon acoustic oscillations (BAO) imprinted in the distribution of low-redshift galaxies. The angular size of the BAO feature, combined with additional data sets to break degeneracies with 
the baryon and matter densities $\omega_{\rm b}$ and $\omega_{\rm m}$, 
can then be used to constrain $H_0$ from large-scale structure with no dependence on the CMB. \citet{abbott2018} combined BAO, big bang nucleosynthesis, and galaxy clustering + weak lensing data from the Dark Energy Survey \citep{des2016} and found $H_0=67.4^{+1.1}_{-1.2}\,{\rm km\,s^{-1}\,Mpc^{-1}}$, consistent with (and independent of) CMB measurements and inconsistent with most local measurements. An identical conclusion was reached using a similar analysis of BAO from the Extended Baryon Oscillation Spectroscopic Survey ($H_0=67.35 \pm 0.97\,{\rm km\,s^{-1}\,Mpc^{-1}}$; \citealt{eboss2020}).

%%%%%%%%%%%%%%%%%%%%%%%%%%%%%%%%%%%%%%%%%%%%%%%%%%%%%%%%%%%%%% 
\begin{table*}
 \renewcommand{\arraystretch}{1.1}
\caption{Cosmological parameters adopted in this paper for the \planck~\citep{planck2020} and EDE \citep{murgia2021} cosmologies. Note that the \planck~analysis adopts $\theta_{\rm MC}$, an approximation to the sound horizon at $z_{\star}$, as one of the six \lcdm~base parameters, and that is what we list in this table. The full numerical solution is reported in \citet{planck2020} as a derived parameter, $100\,\theta_{\star}$. \citet{murgia2021} use $H_0$ rather than $\theta_{\rm MC}$ or $\theta_{\star}$ as a base parameter, so here we list the value of the derived parameter $100\,\theta_{\rm s}(r_{\star})$ for $n=3$ EDE. For each model, we list the marginalized mean and 68\% confidence interval for each parameter, with the best-fit value given in parentheses. The precision we quote is the ratio of the $1\,\sigma$ error --- one half of the 68\% confidence interval --- to the mean value. For $A_{\rm s}$ and $z_{\rm c}$, we quote the precision on the parameter itself as opposed to the precision on the logarithm of the parameter.
}
\label{tab:table1}
\begin{tabularx}{\textwidth}{l >{\centering\arraybackslash}X r >{\centering\arraybackslash}X r}
    \hline
    \hline
    Parameter & \lcdm\ & Precision & $n=3$ EDE & Precision\\
\hline
\rowcolor{Gray}
\textbf{\emph{Base}} & & & &\\
$100\,\theta_{\rm s}(r_{\star})$ & $1.04092 \, (1.040909)\pm0.00031$ & 0.030\,\% 
    & $1.04145\,(1.04106) \pm 0.00037$ & 0.036\,\%\\
$n_{\rm s}$ & $0.9649\,(0.96605) \pm 0.0042$ & 0.44\,\% 
    & $0.9859 \, (0.9844) \pm 0.0069$ & 0.70\,\%\\
$\omega_{\rm b}$ & $0.02237\,(0.022383) \pm 0.00015$ & 0.67\,\% 
    & $0.02281 \, (0.02273)\pm 0.00021$ & 0.94\,\%\\
$\omega_{\rm c}$ & $0.1200\,(0.12011) \pm 0.0012$ & 1.0\,\% 
    & $0.1290 \, (0.1300) \pm 0.0038$ & 3.0\,\%\\
$\ln(10^{10}\,A_{\rm s})$ & $3.044 \,(3.0448) \pm 0.014$ & 1.4\,\% 
    & $3.065\,(3.065) \pm 0.015$ & 1.5\,\%\\
$\tau$ & $0.0544\, (0.0543) \pm 0.0073$ & 13\,\% 
    & $0.0574\,(0.0567) \pm 0.0074$ & 13\,\%\\
\hline
$\Theta_{i}$ & ---& ---& $2.553\,(2.722)\pm 0.56 $ & 22\,\%\\
$\log_{10} z_{\rm c}$ & --- & ---& $3.61\,(3.559) \pm 0.12$ & 28\,\%\\
$f_{\rm EDE}$ & --- & ---& $0.097 \, (0.105) \pm 0.032$ & 33\,\%\\
\hline
\hline
\rowcolor{Gray}
\textbf{\emph{Derived}} & & & &\\
$\tuniv$~[Gyr] & $13.797 \,(13.7971)\pm 0.023$ & 0.17\,\% 
    & $13.246\,(13.210) \pm 0.17$ & 1.3\,\%\\
$r_{\star}$~[Mpc] & $144.43\,(144.39) \pm 0.26$ & 0.18\,\% 
    & $139.43\,(138.95) \pm 1.79$ & 1.3\,\%\\
$H_0\;[{\rm km\,s^{-1}\,Mpc^{-1}}]$ & $67.36\,(67.32)\pm 0.54$& 0.80\,\% 
    & $71.01\,(71.15) \pm 1.05$ & 1.5\,\%\\
$\om$ & $0.3153\,(0.3158)\pm 0.0073$& 2.3\,\% 
    & $0.3022\,(0.3029) \pm 0.0052$ & 1.7\,\%\\
\hline
\end{tabularx}
\end{table*}
%%%%%%%%%%%%%%%%%%%%%%%%%%%%%%%%%%%%%%%%%%%%%%%%%%%%%%%%%%%% 

The strong possibility that the locally-measured value of $H_0$ is meaningfully different from the value inferred from analysis of CMB and large-scale structure data is intriguing: this Hubble tension (see \citealt{di-valentino2021a} for a comprehensive review) points to the prospect of missing physics that modifies the expansion history of the Universe by adding new forms of energy or interactions to the base \lcdm~model (see, e.g., \citealt{knox2020} for a recent overview of classes of solutions). 
It is very difficult to modify only late-time physics and remain consistent with cosmological data sets (see, e.g., \citealt{efstathiou2021}). 
A more promising route for resolving the Hubble tension 
is to posit an expansion rate at early times, prior to the redshift of CMB last scattering ($z_\star=1090$), that is faster than in the standard \lcdm\ model. The sound
horizon at $z_\star$ in such models is smaller than in \lcdm, so the angular diameter
distance between us and $z_\star$ must be reduced in order to maintain the precisely measured angular size of the sound horizon. Such a reduction requires increasing $H_0$. 

A period of ``early dark energy'' (EDE) that, at its peak ($5000 \gtrsim z \gtrsim 3500$), contributes roughly $10\%$ of the total energy density of the Universe before quickly decaying away (\citealt{karwal2016, mortsell2018, agrawal2019, lin2019, poulin2019, sakstein2020, smith2020}, though see \citealt{hill2020} for concerns about EDE's consistency with a variety of data sets) is a tantalizing mechanism for achieving a smaller sound horizon. In this class of models, cosmological ages at $z\ll z_\star$ are only sensitive to the change in the expansion history indirectly, through the accompanying changes in inferred values of $H_0$ and $\om$ based on fits to the CMB. 
Recently-proposed EDE models have best-fit values of $\tuniv \approx 13.0~\gyr$ (e.g., \citealt{klypin2021}), 
meaning the systematic error on $\tuniv$ is at least $\ssim 0.7~\gyr$ (or $\ssim
5\%$). In fact, as we show in this paper, the entire redshift-time relation is
subject to this level of uncertainty ($\ssim 5\%$) as long as cosmological solutions of the Hubble tension related to the pre-recombination expansion rate remain viable. While smaller than the $10\%$ systematic uncertainty that would predict if using the naive $\tuniv \propto H_0^{-1}$ scaling, a 5\% systematic uncertainty is 30 times larger than the error bar quoted by \citet{planck2020} on $\tuniv$ in the base \lcdm\ model and is therefore important to study more closely.

A completely orthogonal handle on the redshift-time relation, and the age of the Universe, comes from the ages of individual stars, stellar remnants, and stellar populations (e.g., metal-poor stars, white dwarfs, and globular clusters and ultra-faint dwarf galaxies) in the local Universe \citep[e.g.,][]{burbidge1957, fowler1960, janes1983, fowler1987, winget1987, cowan1991, cowan1991a, renzini1991, chaboyer1995, vandenberg1996, jimenez1999, krauss2003, verde2013, jimenez2019, verde2019}. 
Stellar ages are independent of cosmological models and, for a known distance and reddening, depend only on the physics of stellar evolution (see \citealt{soderblom2010} for a general review and further discussion in \S~\ref{sec:examples}). Historically, stellar ages were a competitive method to cosmology when determining the age of the Universe (e.g., \citealt{chaboyer1995}, \citealt{vandenberg1996}), but they have
taken a back seat in recent years 
because of the lack of sufficiently precise and accurate distances and uncertainties in some of the underlying stellar physics (e.g., convection, diffusion, opacity, nuclear reaction rates; \citealt{salaris2009, chaboyer2017, arnett2015, dotter2017, valcin2021}) and the dramatic increase in precision in cosmological parameter estimation over the past two decades.

However, as we have just discussed in the context of the Hubble tension, precision cosmology does not necessarily lead to an equally accurate understanding of our Universe and its contents \citep{peebles2002}. In this context, stellar ages have the potential to once again be a useful tool for our understanding of cosmology. Fortuitously, 
observations of stars in the Milky Way (MW) and Local Group (LG) in the midst of a data revolution, and
precise and accurate geometric distances from Gaia \citep{gaia2016}, coupled with time domain insights into stellar interior physics \citep[e.g., K2, TESS][]{howell2014, ricker2015} and precise abundance determinations \citep[e.g., LAMOST, GALAH, APOGEE;][]{cui2012, de-silva2015, majewski2017}, are providing a new foundation for stellar astrophysics. 

In this paper, we explore the uncertainty in ages as derived from cosmological redshifts and consider the role of stellar ages in independently constraining cosmology-derived ages, both now and in the context of projected improvements in near-field observations. 
We first outline the current version of the \lcdm~and an extended version that contains EDE (\S~\ref{sec:cosmology}). We then discuss the translation between redshift and age in cosmology and quantify the uncertainty in this translation that comes from the allowed parameter space in cosmological models, with a particular emphasis on the difference in the redshift-time relation in \lcdm~versus EDE (\S\,\ref{sec:cosmo_ages}). We highlight specific examples of how this uncertainty affects our understanding of galaxy formation and consider the role of stars as cosmology-independent clocks (\S~\ref{sec:examples}).  Finally, we discuss  the age-old issue of accuracy and precision for stellar and cosmological ages and highlight promising areas for improvement (\S~\ref{sec:discussion}). 

%%%%%%%%%%%%%%%%%%%%%%%%%%%%%%%%%%%%%%%%%%%%%%%%%%%%%%%%%%%%%%
\begin{figure*}
 \centering
 \includegraphics[width=0.49\textwidth]{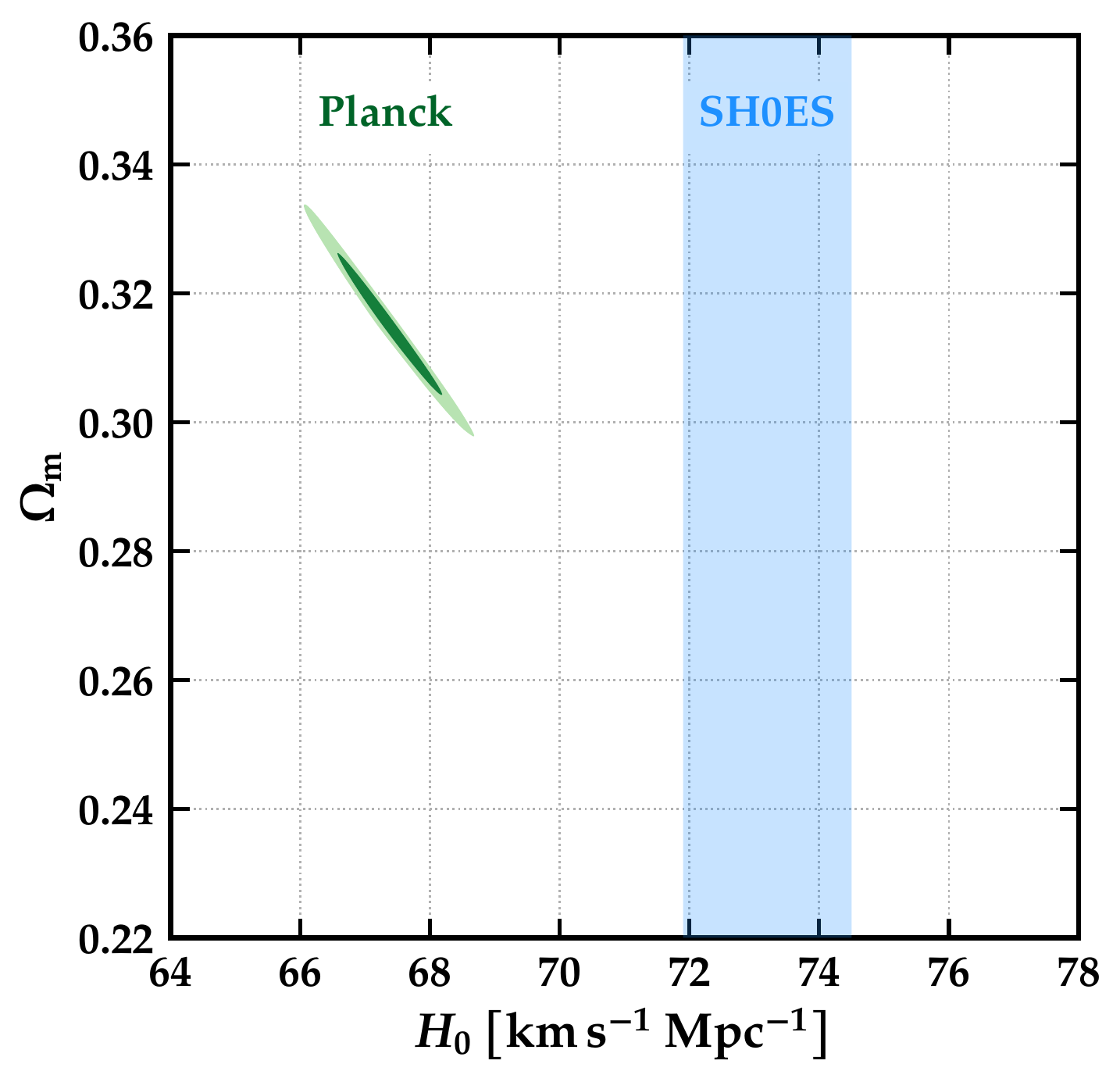}
 \includegraphics[width=0.49\textwidth]{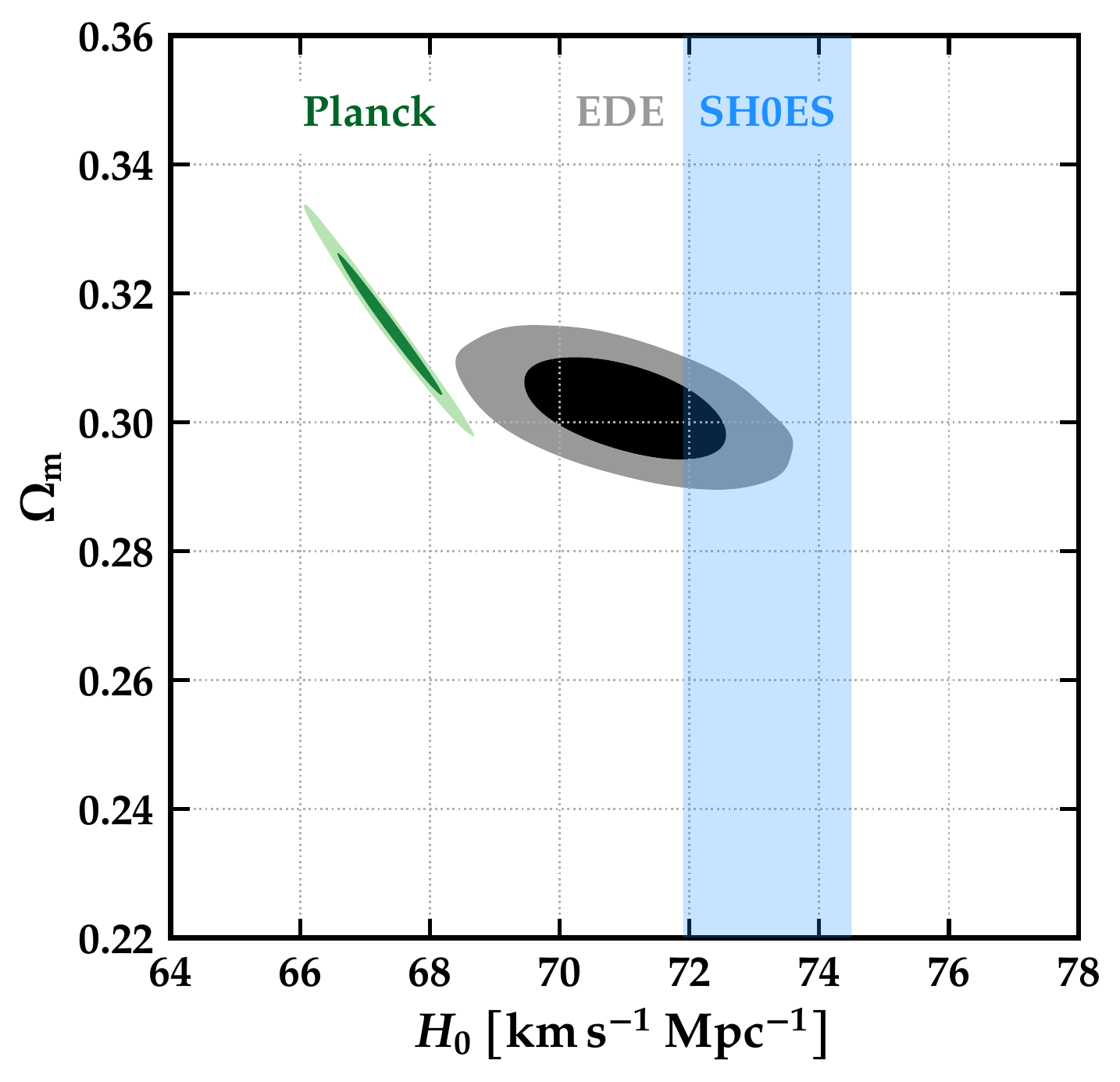}
 \caption{The Hubble tension and its resolution via early dark energy (EDE). \textit{Left:} \planck~constraints on $\om$ and $H_0$ (green contours; the darker contour shows the 68\% confidence interval while the lighter contour shows the 95\% confidence interval), along with the determination of $H_0$ from \citetalias{riess2021} (``SH0ES'') as a representative local value. The Hubble tension is the significant difference in $H_0$ measured locally versus derived from the CMB. \textit{Right:} The resolution of the Hubble tension via EDE. By slightly changing the early expansion history of the Universe relative to the base \lcdm~model, the Hubble tension can be resolved (or at least reduced): the CMB-derived value of $H_0$ increases, and $\om$ decreases, relative to \planck. The $\om-H_0$ degeneracy for the base \lcdm~fit to \planck~is defined by $\om \propto h^{-3}$; for EDE, it is $\om \propto h^{-0.7}$. These correlations are important for understanding behavior in the $\tuniv-H_0$ plane. See Fig.~\ref{fig:t0_h0} and \S\S~\ref{subsec:correlations} and \ref{subsec:ages} for further details.
}
 \label{fig:tension}
\end{figure*}
%%%%%%%%%%%%%%%%%%%%%%%%%%%%%%%%%%%%%%%%%%%%%%%%%%%%%%%%%%%%%% 

\section{Cosmological Preliminaries}
\label{sec:cosmology}
\subsection{The base \texorpdfstring{\bmlcdm}{LCDM}~and EDE cosmologies}
\label{subsec:cosmology}
We adopt the base \lcdm\ model of the \textit{Planck}~analysis \citep{planck2020} as our cosmological standard model. This model, which assumes no spatial curvature and initial conditions that are Gaussian and adiabatic, is 
fully\footnote{By ``fully'', we mean partially. See the text below Eq.~\ref{eq:hub} for additional assumptions that go into the base \lcdm~model.} 
parameterized by 6 numbers: the physical densities of 
CDM and baryons today, $\omega_{\rm c}=\Omega_{\rm c} h^2$ and $\omega_{\rm b}=\Omega_{\rm b} h^2$, where $h$ is the standard dimensionless representation of the present-day Hubble constant, $H_0=100\,h\, \kms\,\mpc^{-1}$, and $\Omega_{\rm X}$ refers to the present-day ratio of the density in component ${\rm X}$ to the critical density $\rho_{\rm crit}=3\,H_0^2/(8\,\pi\,G)$; the amplitude $A_{\rm s}$ and slope $n_{\rm s}$ of the primordial power spectrum of density fluctuations; the angular size of the sound horizon at the epoch of CMB last scattering, $\theta_\star$; and the electron scattering optical depth to reionization, $\tau$~\citep{planck2020}. 
We use best fit, marginalized mean, and 68\% confidence intervals for each parameter based on the {\tt Plik} TT,TE,EE+lowE+lensing likelihood applied to the full-mission data (hereafter, ``\planck'').

In the base \lcdm~model, the expansion rate $H$ at any redshift $z$ or scale factor $a=(1+z)^{-1}$ depends only on 
the current expansion rate, $H_0$, and the physical densities of each energy/matter component --- baryons, CDM, dark energy ($\Lambda$), photons ($\gamma$), and neutrinos ($\nu$) --- as a function of redshift:
\begin{equation}
\label{eq:hub_full}
    H(a)=H_0\,\sqrt{(\Omega_{\rm c}+\Omega_{\rm b})\,a^{-3} + \ol+
      \Omega_{\gamma}\,a^{-4}+\frac{\rho_{\nu}(a)}{\rho_{\rm crit}(a=1)}}\,.
\end{equation}
Furthermore, the base \lcdm~model assumes: (1) the present-day CMB temperature is $T_{\rm CMB}=2.7255\,{\rm K}$ \citep{fixsen2009}, which sets the present-day photon energy density $\rho_{\gamma}$; (2) one massive ($m=0.06\,{\rm eV}$) and two massless neutrino species, which establishes the contribution of neutrinos to the present-day matter density; and (3) the \textit{effective} number of neutrino species\footnote{$N_{\rm eff}$ differs from the (integer) number of neutrino species because neutrino decoupling is not complete at the time of the electron-positron annihilation in the early Universe. Since neutrinos have energy-dependent interactions with the photon-baryon plasma before and during decoupling, with higher-energy neutrinos interacting more strongly, the energy spectrum of the neutrinos is distorted slightly relative to the assumed Fermi-Dirac distribution. This spectral distortion, and the accompanying slight decrease in the difference between the neutrino and photon energy densities relative to the assumption of a thermal neutrino spectrum, can be accounted for by using $N_{\rm eff}=3.046$ rather than 3 \citep{mangano2005}.} is $N_{\rm eff}=3.046$, which provides the conversion from $\rho_{\gamma}$ to $\rho_{\rm r}$ at early times, when all three neutrino species are relativistic: $\rho_{\rm r}(z \gtrsim 100)=1.692 \,\rho_{\gamma}(z \gtrsim 100)$.
The radiation density $\omega_{\rm r}$ is therefore fixed --- $\orad=4.1837\times10^{-5}\,h^{-2}$ --- and, to a good approximation, the expansion history depends only on the parameters $H_0$ and $\om$:
\begin{equation}
\label{eq:hub}
    H(a)=H_0\,\sqrt{\om\,a^{-3} + (1-\om)+\orad\,a^{-4}}\,.
\end{equation}
Accordingly, we will frequently consider the $\om-H_0$ plane in what follows (see also \citealt{lin2020}).

As a fiducial model that is consistent with both the CMB and with the SH0ES value of $H_0$, we adopt the EDE model described in \citet[hereafter \citetalias{poulin2018}; see also \citealt{smith2020} and \citealt{murgia2021}]{poulin2018},
in which EDE is characterized by three quantities: the fractional contribution of EDE to the energy density of the Universe ($f_{\rm EDE}$) at the redshift ($z_{\rm c}$) where the EDE field becomes dynamical and the initial value of the field ($\Theta_{i}$). In a more general model, the power law exponent $n$ of the EDE potential --- which is related to the asymptotic equation-of-state parameter for EDE --- can also vary; we restrict our analysis to the $n=3$ case, which is generally very close to the best-fit for the more general case. 

Our EDE results are based on parameters derived from the same \textit{Planck} 2018 analysis of  TT,TE,EE+lowE+lowL+lensing plus BAO+SNIa+SH0ES+FS (see Table~1 of \citealt{murgia2021} and accompanying discussion). The SH0ES constraint used in \citet{murgia2021} is from \citet{riess2019} and is slightly larger in terms of both its central value and error --- $H_0=74.03 \pm 1.42\,{\rm km\,s^{-1}\,Mpc{^{-1}}}$ --- than the more recent \citet{riess2021} result of $H_0=73.2 \pm 1.3\,{\rm km\,s^{-1}\,Mpc{^{-1}}}$ that we adopt as a representative local value in this work. This difference does not affect our results qualitatively and should have at most a very minor quantitative effect. Table~\ref{tab:table1} contains the mean and $\pm1\,\sigma$ values, as well as the best fit value, for each parameter in both models. We adopt EDE values taken directly from the MCMC output, which differ slightly (but unimportantly, for our purposes) from those obtained with a further minimization algorithm, as was done for the best-fit values presented in \citet{murgia2021}. We also list the precision --- the ratio of the $1\,\sigma$ error to the mean value --- of the measurement for each parameter for both \planck~and EDE. Appendix~\ref{sec:appendix} contains a brief exploration how EDE affects the cosmological expansion history and derived value of $H_0$.

Figure~\ref{fig:tension} shows the basic Hubble tension (left panel) and its resolution via EDE (right panel). In the base \lcdm\ model, \planck\ results provide a tight constraint in $\om-H_0$ parameter space. While local measurements of $H_0$ do not provide information about $\om$, the locally-measured value of $H_0$ is sufficiently large that no value of $\om$ provides consistency with \planck. By extending the base model to include a different expansion history --- in this case, a period of EDE --- agreement between CMB and local measurements of $H_0$ can be obtained (right panel of Fig.~\ref{fig:tension}), as EDE pushes the preferred value of $\om$ slightly lower and the value of $H_0$ somewhat higher (with a non-trivially larger error). 

As is clear from Figure~\ref{fig:tension}, CMB observations impose a tight connection between $\om$ and $H_0$ for the base \lcdm~model. A heuristic explanation of this correlation is useful in understanding the effects of EDE on cosmological quantities, including ages.

\subsection{Parameter correlations in the base \texorpdfstring{\bmlcdm}{LCDM}~model}
\label{subsec:correlations}
The best-constrained cosmological parameter 
is $\theta_{\star}$, which sets the acoustic scale for oscillations in the photon-baryon fluid and therefore can be determined by the spacings of the peaks in the CMB power spectrum. $\theta_{\star}$ is the ratio of the sound horizon at last scattering, $r_{\star}$, to the angular diameter distance to the last scattering surface, $d_{\star}$. Roughly speaking, then, $r_{\star}$ is sensitive to 
pre-recombination physics while $d_{\star}$ is sensitive to 
post-recombination physics, and any change in pre-$z_{\star}$ physics relative to the base \lcdm~model must be balanced by an accompanying post-$z_{\star}$ change (and vice versa) to keep $\theta_\star$ fixed.

The comoving sound horizon at $z_{\star}$ is 
\begin{equation}
\label{eq:rsound}   
    r_{\star} = r_{\rm s}(a_{\star})=\int_0^{a_{\star}} \frac{c_{\rm s}(a)}{a^2\,H(a)}\,da\,.
\end{equation}
The sound speed, $c_{\rm s}(a) \approx c/\sqrt{3}$ it varies with redshift as 
\begin{equation}
    c_{\rm s}(a)=\frac{c}{\sqrt{3}}\,\frac{1}{\sqrt{1+R(a)}}\,,
\end{equation}
where $R(a)=\frac{3}{4}\,\rho_{\rm b}(a)/\rho_{\gamma}(a)=R_0\,a$ is the redshift-dependent ratio of momenta in baryons to photons. Integrating Eq.~\ref{eq:rsound}, the sound horizon is equal to 
\begin{flalign}
    r_{\star} = 2\frac{c}{H_0}\frac{1}{\sqrt{3\,\om\,R_0}}\,
    \Bigg[ &\arcsinh\left(\sqrt{\frac{a_{\star}/a_{\rm eq}+1}{1/R_{\rm eq}-1}}\right) \nonumber \\ -&\arcsinh\left(\sqrt{\frac{1}{1/R_{\rm eq}-1}}\right)\Bigg]\,
    \label{eq:rsound_analytic}
\end{flalign}
 (see also \citealt{hu1995}). This equation depends on $\om$ and $h$ both explicitly and implicitly via the scale factor at matter-radiation equality, $a_{\rm eq} \propto (\om\,h^2)^{-1}$, and $R_{\rm eq} = R_0\,a_{\rm eq}$. Dependence on the baryon density $\omega_{\rm b}$ comes through $R_0$ (and therefore $R_{\rm eq}$). The integral in Eq.~\ref{eq:rsound} lies entirely in the epoch where the single $m>0$ Standard Model neutrino is relativistic ($m_{\nu}=0.06\,{\rm eV}$ in the base \lcdm\ model, and $3.15\,kT_{\nu}(a_{\star})=0.58\,{\rm eV}$), meaning that the neutrino contribution must be removed from $\om$ (both explicitly and in computing $a_{\rm eq}$) when evaluating Eq.~\ref{eq:rsound_analytic} if high precision is required.
The dependence of $r_{\star}$ on $\om$ and $H_0$ near ($\Omega_{\rm m, Pl},\,h_{\rm Pl}$) is 
\begin{equation}
\label{eq:rstar_cosmo}
    \frac{r_{\star}}{r_{\star, {\rm Pl}}} = \left(\frac{\om}{\Omega_{\rm m, Pl}}\right)^{-0.25}\left( \frac{h}{h_{\rm Pl}}\right)^{-0.49}
    \approx \,\left(\frac{\omega_{\rm m}}{\omega_{\rm m, Pl}}\right)^{-1/4}\,.
\end{equation}
We fix the baryon density to the \planck\ value in deriving Eq.~\ref{eq:rstar_cosmo}; we find  
$r_{\star}\propto \omega_{\rm b}^{-0.096}$ near the \planck~fit to the base \lcdm~model at fixed $\om$ and $h$.

The comoving angular diameter distance to $a_{\star}$ is equal to 
\begin{flalign}
\label{eq:dstar}
        d_{\star} &= \int_{a_{\star}}^1 \frac{c}{a^2\,H(a)}\,da\,\\
&=\frac{c}{H_0}
  \,\int_{a_{\star}}^1 \left[\om\,a+\ol\,a^4 + \orad \right]^{-1/2}\,da\,.
\end{flalign}
Although there is no analytic expression for $d_{\star}$ even in the base \lcdm\ model, we can obtain an analytic approximation that is accurate to essentially arbitrary precision by splitting the integral in Eq.~\ref{eq:dstar} into one portion where $\Omega_{\Lambda}$ is negligible and another where $\Omega_{\rm r}$ is negligible:
\begin{flalign}
\label{eq:dstar_split}
         d_{\star}=\frac{c}{H_0} \bigg(
  &\int_{a_{\star}}^{a_{i}} \left[\om\,a+\orad \right]^{-1/2}\,da \;\nonumber\\
  +\,&\int_{a_{i}}^1 \left[\om\,a+\ol\,a^4\right]^{-1/2}\,da\,
  \bigg)\,.
\end{flalign}
The first integral in Eq.~\ref{eq:dstar_split} is straightforward, while the second can be expressed in terms of Gauss' hypergeometric function $_2F_1(\nicefrac{1}{2}, \nicefrac{1}{6}, \nicefrac{7}{6}; Z)$, with $Z=a^3\,(\om-1)/\om$. A natural choice for $a_{i}$ in Eq.~\ref{eq:dstar_split} is the scale factor of $\Lambda$-radiation equality, $a_{\Lambda {\rm eq}}=(\orad/\ol)^{1/4} \approx 0.1$, giving
\begin{equation}
\label{eq:dstar_approx}
    d_{\star}=\frac{2\,c}{H_0\,\om^{1/2}}\,\left(\sqrt{a+a_{\rm eq}}\,\bigg|_{a_{\star}}^{a_{\Lambda {\rm eq}}} + \sqrt{a}\, {_2}F_1(\nicefrac{1}{2}, \nicefrac{1}{6}, \nicefrac{7}{6}; Z)\,\bigg|_{a_{\Lambda {\rm eq}}}^1\right)\,.
\end{equation}
The cosmological parameter dependence of $d_{\star}$ is 
\begin{equation}
\label{eq:dstar_cosmo}
    \frac{d_{\star}}{d_{\star, {\rm pl}}} = \left(\frac{\om}{\Omega_{\rm m, Pl}}\right)^{-0.4}\left( \frac{h}{h_{\rm Pl}}\right)^{-1}\,
\end{equation}
(see also \citealt{vittorio1985}).

 The sound horizon at last scattering, $\theta_{\star}=r_{\star}/d_{\star}$, obtained using Equations~\ref{eq:rsound_analytic} and \ref{eq:dstar_approx} is  identical to the derived \planck~value to 0.002\% (the quoted error on the \planck~value is 0.03\%). 
 Combining Eqs.~\ref{eq:rstar_cosmo} and~\ref{eq:dstar_cosmo}, we find that 
\begin{equation}
\label{eq:theta_cosmo}  
    \frac{\theta_{\star}}{\theta_{\star, {\rm pl}}} = \left(\frac{\om}{\Omega_{\rm m, Pl}}\right)^{0.15}\left( \frac{h}{h_{\rm Pl}}\right)^{0.51}\,.
\end{equation}
It is therefore this combination of cosmological parameters in the base \lcdm~model --- $\om^{0.15}\,h^{0.51}$, or~$\om\,h^{3.4}$ --- that is well-constrained by the highly precise measurement of $\theta_{\star}$ and is the origin the very narrow \planck~confidence contours in Fig.~\ref{fig:tension}. The orientation of the degeneracy in $\om-H_0$ space is actually slightly different from a curve of constant $\theta_{\star}$, as the measured values of $\om$ and $H_0$ depend on information from the heights of the peaks (which are mostly sensitive to $\om\,h^2$) in addition to their spacings. In practice, this shift relative to the $\om\,h^{3.4}$ degeneracy is relatively small, to $\om\,h^{3}\approx\mathrm{constant}$ \citep{percival2002, kable2019, planck2020}.  

EDE contributes non-negligibly to the energy density, and therefore to the expansion rate $H(a)$, at early times but not late times; accordingly, the expressions above for $r_{\star}$ (and the parameter dependence of $\theta_{\star}$) need to be modified in the presence of EDE but those for $d_{\star}$ do not. Appendix~\ref{sec:appendix} discusses $H(a)$ for the EDE model considered here. The global effect of these changes is to modify the EDE confidence contour in $\om-H_0$ space: the contour is much shallower (less change in $\om$ as $H_0$ is varied) --- approximately defined by a constant value of $\om \, h^{0.7}$ --- and much broader than the contour defined by the \planck~fit to the base \lcdm~model.

\section{Cosmological ages and times}
\label{sec:cosmo_ages}
\subsection{From scale factor to time and back again}
\label{subsec:a-t}

%%%%%%%%%%%%%%%%%%%%%%%%%%%%%%%%%%%%%%%%%%%%%%%%%%%%%%%%%%%%%%
\begin{figure*}
 \centering
 \includegraphics[width=0.50\textwidth]{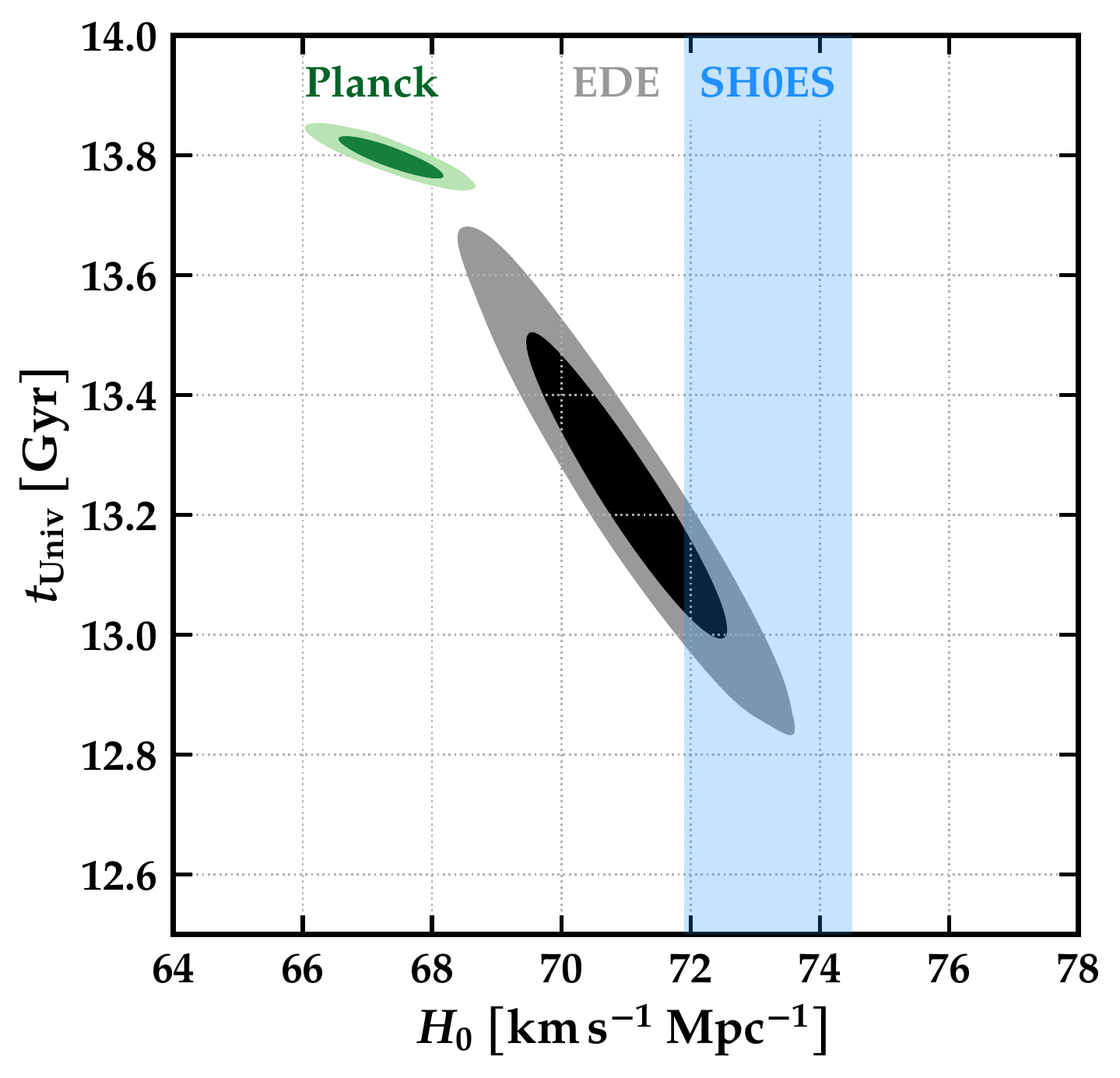}
  \includegraphics[width=0.494\textwidth]{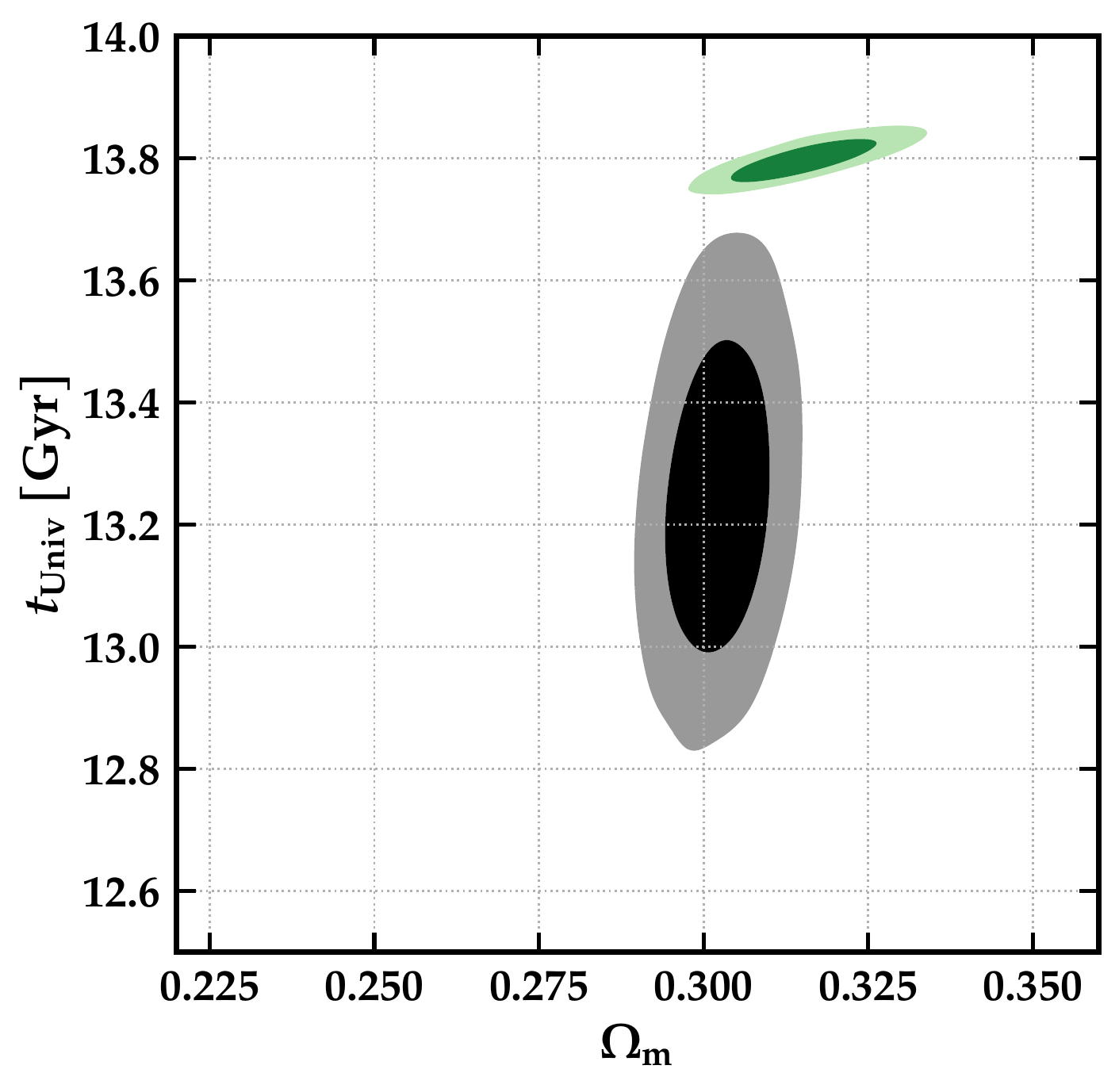}
\caption{\textit{Left}: The Hubble tension using age rather than $\om$ as a secondary parameter (compare with Fig.~\ref{fig:tension}). The portion of EDE parameter space that is consistent with the \citetalias{riess2021} value of $H_0$ results in a value of $\tuniv\ssim 13.0-13.2\,\gyr$ that is significantly smaller than in the base \planck~model ($13.8\,\gyr$). Because of the parameter correlations in the base \lcdm~model enforced by the CMB acoustic peaks, $\om\,h^{3}\propto {\rm constant}$, the \planck~confidence contours in the $\tuniv-h$ plane are $\tuniv \propto h^{-0.2}$. This differs substantially from the naive scaling of $\tuniv \propto h^{-1}$. The EDE contours, on the other hand, are defined by $\tuniv \propto h^{-0.9}$,  
 nearly the same as the naive scaling, because the parameter combination $\om\,h^{0.7}$ is best constrained in EDE.
 See \S~\ref{subsec:ages} for details. \textit{Right:} The dependence of $\tuniv$ on $\om$ in the base \planck~model (green) and EDE (black and gray). While $\tuniv$ is precisely determined and only depends very weakly on $\om$ in the base \lcdm~model, $\tuniv$ is much less well determined in EDE (though $\om$ is better constrained to a slightly lower value than in base~\lcdm).}
 \label{fig:t0_h0}
\end{figure*}
%%%%%%%%%%%%%%%%%%%%%%%%%%%%%%%%%%%%%%%%%%%%%%%%%%%%%%%%%%%%%% 

 This age of the Universe in the base \lcdm\ model at any scale factor $a$ is calculated via a 
straightforward integral: 
\begin{flalign}
  \label{eq:t_int}
  t(a)&=\int_0^a \frac{1}{a\,H(a)}\,da\\
  &=\frac{1}{H_0}
  \,\int_0^a\left[\om\,a^{-1}+(1-\om)\,a^2 + \orad\,a^{-2}\right]^{-1/2}\,da\,.
\end{flalign}
For scale factors well into the matter-dominated era, $a \gg \orad/\om \approx 2.9\times 10^{-4}$, the radiation density term
can be safely ignored for calculations requiring percent level accuracy and the
resulting integral is analytic:
defining
\begin{equation}
  \label{eq:ttilde}
  \ttilde \equiv \frac{2}{3}\frac{1}{
H_0\,\sqrt{1-\om}} =6.519\,\frac{1}{\sqrt{1-\om}}\,h^{-1}\,\gyr\,,
\end{equation}
the time-scale factor relation is 
\begin{equation}
  \label{eq:t_of_a}
  t(a) = \ttilde\,\arcsinh\left(\sqrt{\frac{1}{\om}-1}\,a^{3/2}\right)\,,
\end{equation}
while the inverse relation is 
\begin{equation}
    \label{eq:a_of_t}
    a(t)=\left[\sqrt{\frac{\om}{1-\om}}\,\sinh{\left(\frac{t}{\ttilde}\right)}  \right]^{2/3}\,.
\end{equation}
The age of the Universe is then 
\begin{flalign}
  \label{eq:tuniv}
  \tuniv \equiv t(a=1) &=
  \ttilde\,\arcsinh\left(\sqrt{\frac{1}{\om}-1}\right)\\
  &=\ttilde\,\ln\left(\sqrt{\frac{1}{\om}-1}+\sqrt{\frac{1}{\om}}\right) \label{eq:tuniv_alt}
\end{flalign}
and is equal to $13.80\,\gyr$ when adopting the Planck
parameters. The cosmological parameter dependence of Eq.~\ref{eq:tuniv} near the \planck~values can be approximated as
\begin{equation}
  \label{eq:tuniv_cosmo}
  \frac{\tuniv}{t_{\rm Pl}} =\left(\frac{\Omega_{\rm
            m}}{\Omega_{\rm m, Pl}}\right)^{-0.28}\, \left(\frac{h}{h_{\rm Pl}}\right)^{-1}\,
\end{equation}
or $\tuniv \propto \om^{-0.28}\,h^{-1}\propto \omega_{\rm m}^{-0.28}\,h^{-0.44}$; This relationship holds for EDE models (constrained by \textit{Planck} data) as well and reinforces that the age of the Universe is sensitive to both $\om$ and $H_0$ in flat \lcdm-like cosmological models. 
Comparing Eq.~\ref{eq:tuniv_cosmo} with Eq.~\ref{eq:theta_cosmo}, we see that the cosmological dependence of $\theta_{\star}$ and $\tuniv$ are closely related: holding all parameters except $\om$ and $h$ fixed, \planck~requires
\begin{equation}
\label{eq:theta-t-corr}
    \tuniv \propto \theta_{\star}^{-2}
\end{equation}
for the base \lcdm\ model. Marginalizing over the additional parameters --- of which $\omega_{\rm b}$ is the most important, given its role in setting $r_{\star}$ --- modifies Eq.~\ref{eq:theta-t-corr} to $\tuniv \propto \theta_{\star}^{-5.5}$. This accidental correlation between $\tuniv$ and $\theta_{\star}$ explains why the \planck~constraint on $\tuniv$ is so good ($0.17\%$) --- more precise than all of the primary parameters except $\theta_{\star}$ --- using the base \lcdm\ fit even though $\tuniv$ depends on the much less precisely determined parameters $\om$ and $H_0$ (see also \citealt{hu2001,knox2001}). EDE does not have the same $\theta_{\star}-\tuniv$ correlation: although $\theta_{\star}$ is precisely determined, it is not tightly connected to $\tuniv$ in EDE, which is reflected in the  much larger error bar on $\tuniv$ for EDE than for \planck\ in Table~\ref{tab:table1}.

%%%%%%%%%%%%%%%%%%%%%%%%%%%%%%%%%%%%%%%%%%%%%%%%%%%%%%%%%%%%%%
\begin{figure*}
 \centering
 \includegraphics[scale=0.54]{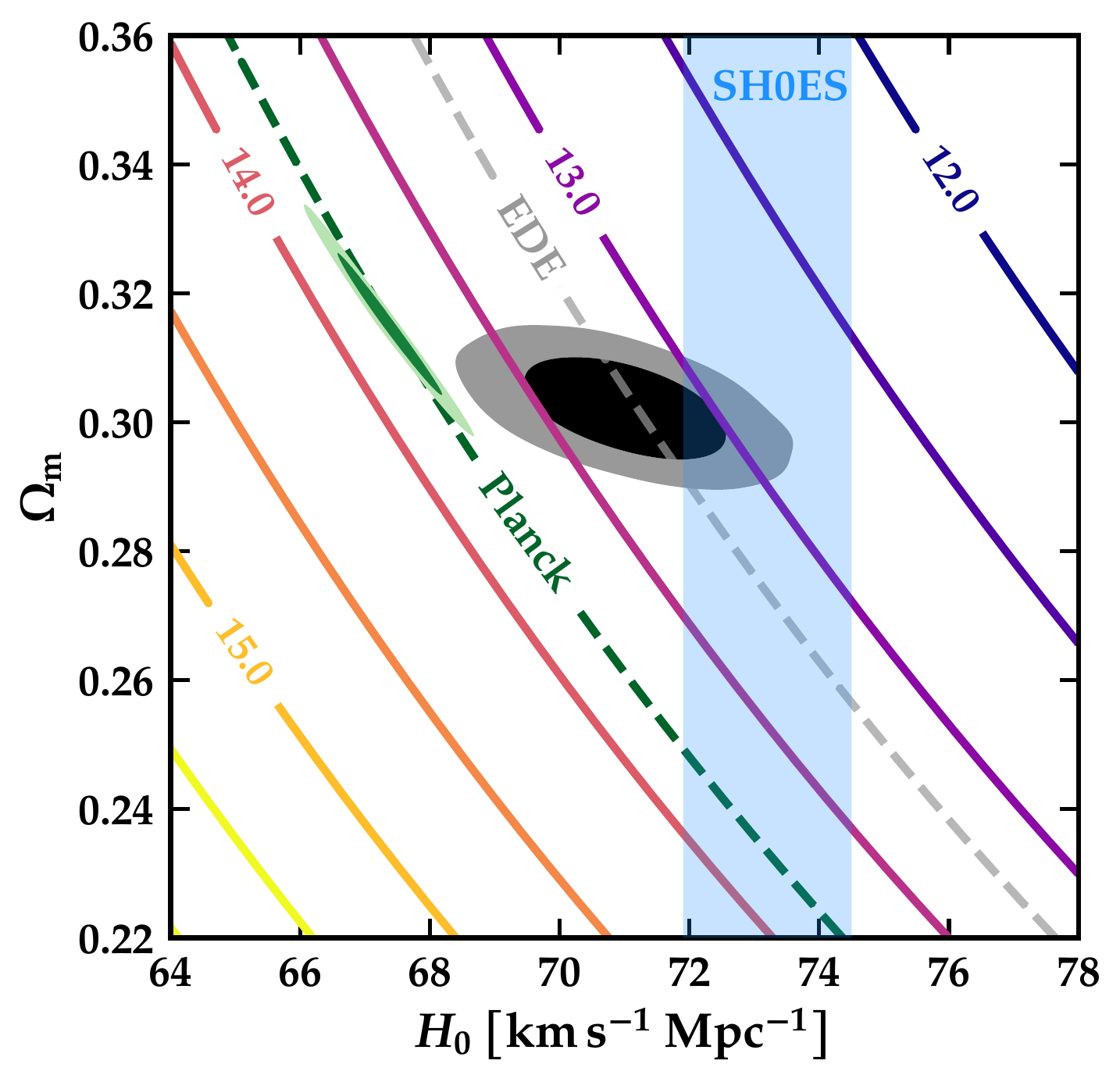}
 \includegraphics[scale=0.54]{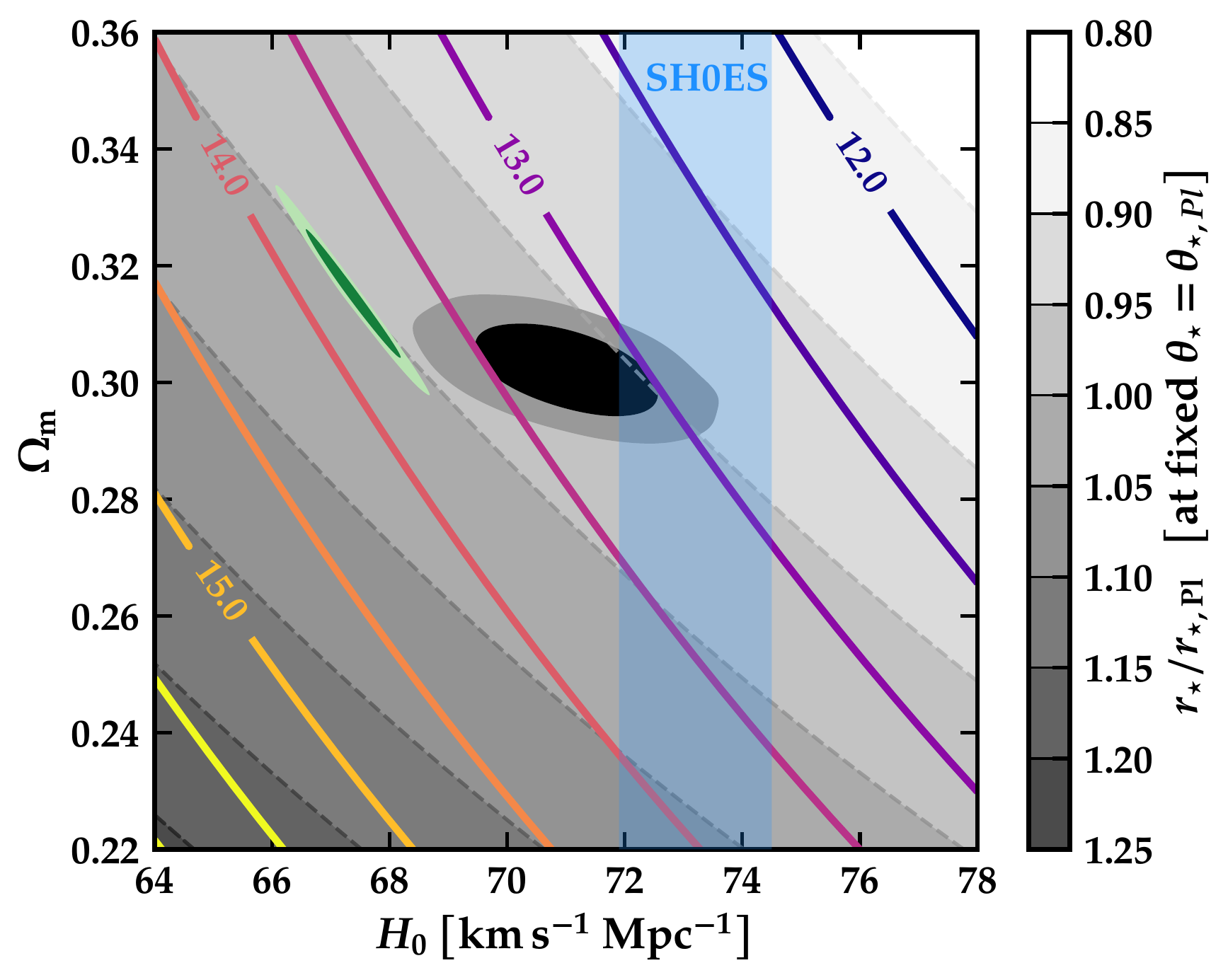}
 \caption{\textit{Left:} Contours of constant age (in Gyr) in $(\om,\,H_0)$ parameter space. As in Figs.~\ref{fig:tension} and \ref{fig:t0_h0}, the $1\,\sigma$ and $2\,\sigma$ 2D confidence contours for \textit{Planck}~fits to the base \lcdm\ model are shown in dark and light green, with the green dashed line showing the age of the Universe --- $\tuniv=13.80~\gyr$ --- for this model.
 Equivalent 2D constraints for a representative EDE model \citep{murgia2021} are plotted in black and gray, with the gray dashed line showing the corresponding age of the Universe, $\tuniv=13.21~\gyr$. The SH0ES measurement of $H_0$ \citepalias{riess2021} is shown as a blue shaded vertical band with width that encompasses the $\pm\,1\,\sigma$ error. 
 \textit{Right:} Same as left, with gray-scale filled contours that show how $d_{\star}$ and $r_{\star}$ change, in intervals of $5\%$, if $\theta_{\star}$ is fixed to the \planck~value (with darker shading indicating increasing values of $r_{\star}$)
 Models that resolve the Hubble tension by reducing $r_{\star}$ by ${\sim}4\%$, such as EDE, require lower values of $\om$, higher values of $H_0$, and lower values of $\tuniv$ than the base \lcdm\ fit to \textit{Planck}~data. Note that the \planck~age contour aligns well with the 2D confidence contours for \planck, which is a result of $\theta_{\star}$ and $\tuniv$ having nearly the same scaling for the base \lcdm~model. The same is not true for EDE. See \S~\ref{subsec:correlations} for details.
 \label{fig:h0_om_t0}
}
\end{figure*}
%%%%%%%%%%%%%%%%%%%%%%%%%%%%%%%%%%%%%%%%%%%%%%%%%%%%%%%%%%%%%% 

At early times, Eq.~\ref{eq:t_of_a} can be
expressed as
\begin{equation}
  \label{eq:t_early}
  t(a) \cong \frac{2}{3}\omega_{\rm m}^{-1/2}\,a^{3/2}\;\;\; (a_{\rm eq} \ll a \lesssim 0.25)\,,
\end{equation}
which makes it clear that 
$t(a)$ is only sensitive to the physical matter density of the Universe $\omega_{\rm m}(a)$ for both the base
\lcdm\ model and EDE in the heart of the matter-dominated era $(100 \gtrsim z \gtrsim 3$). Near the present day, Eq.~\ref{eq:t_of_a} is approximately 
\begin{equation}
\label{eq:t_late}
    t(a) \cong \tuniv-\frac{1-a}{H_0}=\tuniv-\frac{1}{H_0}\frac{z}{1+z} \;\;\; (a\approx 1)\,.
\end{equation}

It is useful in many astrophysical settings to consider times relative to the present day rather than the beginning of the Universe. This \textbf{lookback time} ($t_{\rm lb}$) is simply 
\begin{equation}
    \label{eq:tlb}
    \tlb(a) \equiv \tuniv-t(a)\,.
\end{equation}
The lookback time as a function of scale factor is 
\begin{equation}
\label{eq:tlb_a} 
\tlb(a)=\ttilde\,\left[\arcsinh\left(\sqrt{\frac{1}{\om}-1}\right) - \arcsinh\left(\sqrt{\frac{1}{\om}-1}\,a^{3/2}\right)\right]
\end{equation}
Equation~\ref{eq:tlb_a} can be inverted to give the scale factor as a function of lookback time:
\begin{equation}
  \label{eq:a_of_tlb}
  a(\tlb)=\left[\cosh\left(\frac{\tlb}{\ttilde}\right)-\sqrt{\frac{1}{1-\om}}\,\sinh\left(\frac{\tlb}{\ttilde}\right)\right]^{2/3}\,.
 \end{equation}
 The expressions for $t(a), \,\tuniv$, and $\tlb(a)$ derived in this section apply for both the base \lcdm~model and EDE for $a \gtrsim 10\,a_{\star}$ (i.e., $z \lesssim 100$): for this range of scale factors, the integral in Eq.~\ref{eq:t_int} is dominated by epochs where the contributions of EDE are negligible. At earlier times ($a \lesssim 10\, a_{\star}$), calculations of $t(a)$ must be modified to directly include the effects of EDE on the expansion rate; see Appendix~\ref{sec:appendix} for the appropriate functional form of $H(a)$ for the EDE model considered in this paper.
 
\subsection{Ages in \planck~and EDE cosmologies}
\label{subsec:ages}
%%%%%%%%%%%%%%%%%%%%%%%%%%%%%%%%%%%%%%%%%%%%%%%%%%%%%%%%%%%%%%
\begin{figure}
 \centering
 \includegraphics[width=\columnwidth]{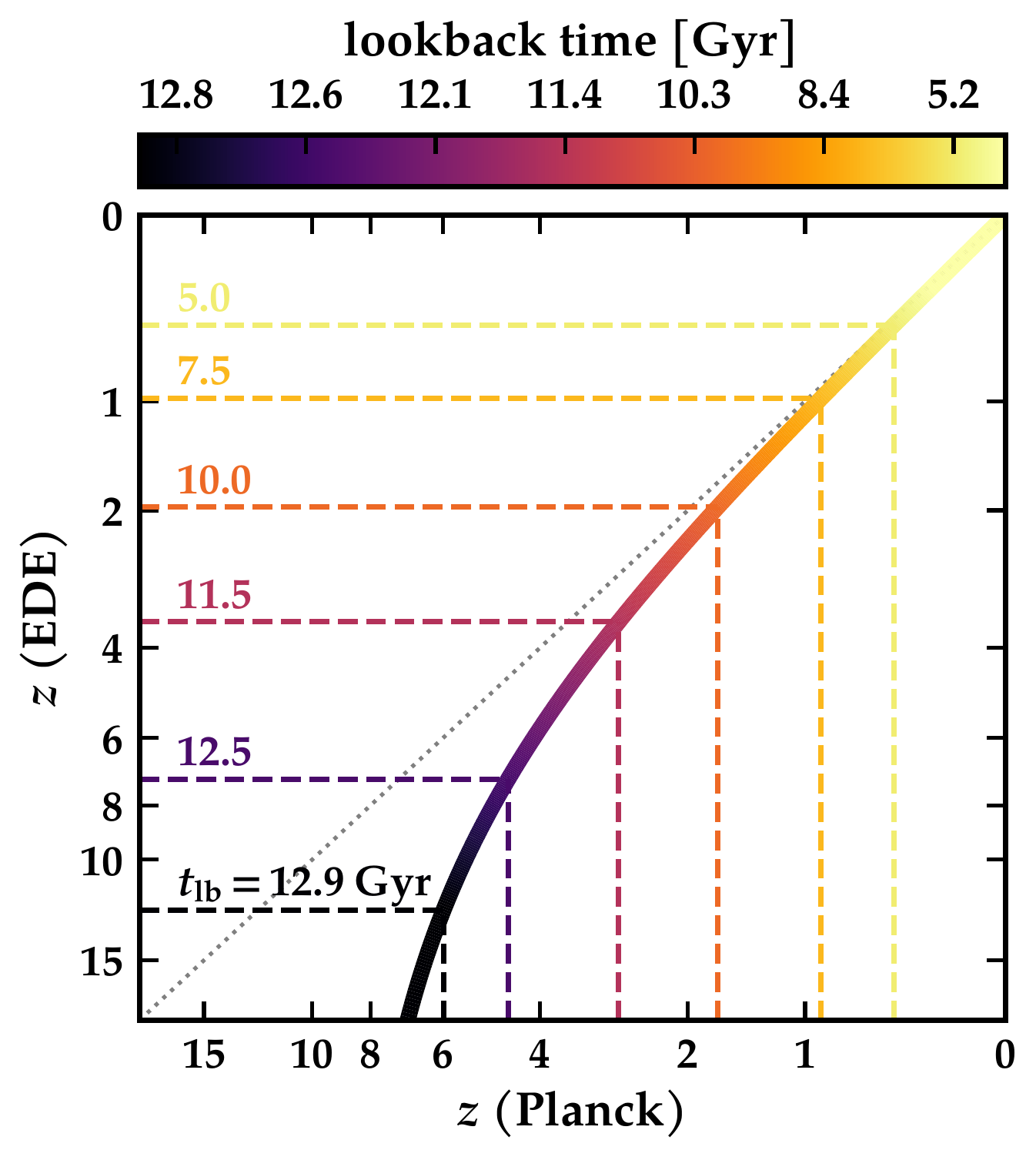}
 \caption{The redshift corresponding to a given lookback time in the 
   Planck ($x$-axis) and EDE ($y$-axis) cosmologies. Since the cosmological parameters differ in the two models, the redshift corresponding to a fixed $\tlb$ differs as well. 
   The difference in $z(t_{\rm lb})$ is minimal for $\tlb \lesssim 8\,\gyr$ but becomes substantial for larger values of $\tlb$: $z_{\rm EDE}=8$
   corresponds to the same lookback time (12.60 Gyr) as  $z_{\rm Pl}=4.89$, while a lookback time of 12.87 Gyr occurs at $z_{\rm EDE}=12.35$ versus $z_{\rm Pl}=6$. 
   The age of the Universe in the EDE model, $13.21\,\gyr$, corresponds to $z_{\rm Pl}=8.5$.  
 \label{fig:z_vs_z}
}
\end{figure}
%%%%%%%%%%%%%%%%%%%%%%%%%%%%%%%%%%%%%%%%%%%%%%%%%%%%%%%%%%%%%% 

%%%%%%%%%%%%%%%%%%%%%%%%%%%%%%%%%%%%%%%%%%%%%%%%%%%%%%%%%%%%%
\begin{figure}
 \centering
 \includegraphics[width=\columnwidth]{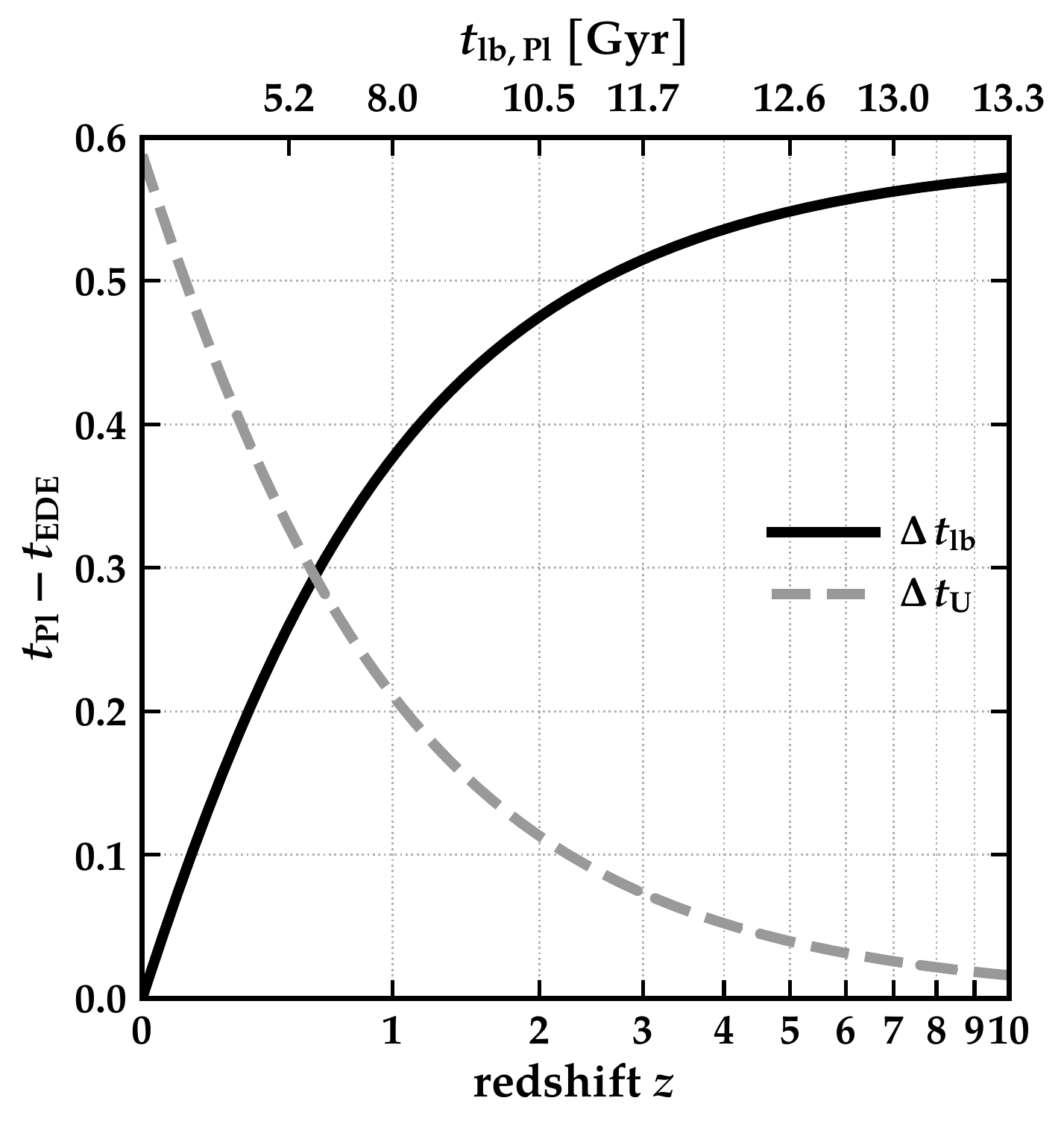}
 \caption{The difference in age as a function of redshift between the \planck\ and EDE models in terms of cosmic time (gray) and lookback time (black). The difference between cosmic times in the two models is lowest (in absolute terms) at high redshift, while the difference between lookback times is lowest at low redshift. The difference is bounded by the difference in the age of the Universe in the two models, $t_{\rm 0,Pl}-t_{\rm 0, EDE} \approx 0.6\,\gyr$.
 \label{fig:tlb_tuniv}
}
\end{figure}
%%%%%%%%%%%%%%%%%%%%%%%%%%%%%%%%%%%%%%%%%%%%%%%%%%%%%%%%%%%%%% 

With these calculations of cosmological ages in hand, we can revisit the Hubble tension in $\tuniv-H_0$ space. 
The left panel of Figure~\ref{fig:t0_h0} shows that \planck~gives a very precise value of $\tuniv$ and that local measurements of $H_0$ are agnostic as to the age of the Universe because they are not directly sensitive to $\om$. Resolving the Hubble tension by introducing EDE results in a Universe that is non-trivially younger than the \planck~cosmology. Since EDE changes the orientation of the best-fit confidence contours in $\om-H_0$ space relative to the base \lcdm~contour (see Fig.~\ref{fig:tension} and discussion in \S~\ref{subsec:correlations}), the contours also align differently in $\tuniv-H_0$ space. Using Eq.~\ref{eq:tuniv_cosmo} and the fact that the degeneracy in $\om-H_0$ space is approximately defined by a constant value of $\theta_{\star}$ for the base \lcdm~model, we find that $\tuniv \propto h^{-0.2}$ for \planck~(because $\om\propto h^{-3}$). The $\om-H_0$ degeneracy in EDE is much broader and shallower ($\om\propto h^{-0.7}$;~c.f.~the black/gray contours in the right panel of Fig.~\ref{fig:tension}), which results in $\tuniv \propto h^{-0.9}$, close to the naive scaling of $\tuniv \propto H_0^{-1}$.

The right panel of Fig.~\ref{fig:t0_h0} shows $\tuniv-\om$ parameter space. While \planck~constrains $\tuniv$ much better than $\om$ (0.17\% versus 2.3\% precision), constraints on the EDE model result in roughly similar precision for the two parameters (1.3\% versus 1.7\%). The age of the Universe covers a much wider portion of the EDE parameter space, and $\om$ covers a somewhat narrower portion, compared to \planck. As noted above, this difference has its origins in the high-precision, and nearly cosmological-model-independent, determination of $\theta_{\star}$ from \textit{Planck} data and the dependence of $\theta_{\star}$ (and the acoustic peak heights) on $(\om,H_0)$ in the two models.

One way to encompass all of the relevant information is to consider $(\om,H_0)$ parameter space once again. Since $\tuniv$ depends only on these two parameters, it is possible to draw contours of constant $\tuniv$ in this space. The left panel of Figure~\ref{fig:h0_om_t0} shows this parameter space, with contours of constant $\tuniv$ (in Gyr) labeled. Cosmological constraints on the parameters are shown for \planck~(green) and EDE (black/gray), and the local value of $H_0$ is shown in light blue, as before.
The figure shows the tight correlation between the $\om-H_0$ degeneracy and the contour of constant age for the \planck~cosmology. The degeneracy for the EDE cosmology is both broader and less well aligned with the age contours. We can also use this parameter space to get an intuition about the effects of EDE by fixing $\theta_\star$ to the \planck~value and varying $d_{\star}$, which depends only on $\om$ and $H_0$, by intervals of 5\%; to keep $\theta_\star$ fixed, this also requires changing $r_\star$ by 5\% intervals. This effect is shown as the gray-scale contours in the right panel of Fig~\ref{fig:h0_om_t0}. 
To resolve the Hubble tension, EDE reduces $r_{\star}$ by ${\sim}4\%$ relative to \planck, which requires a higher value of $H_0$ and a slightly lower value of $\om$ and results in a lower value of $\tuniv$.

\subsection{The redshift-age relation}
\label{subsec:age_t_relation}
%%%%%%%%%%%%%%%%%%%%%%%%%%%%%%%%%%%%%%%%%%%%%%%%%%%%%%%%%%%%%
\begin{figure*}
 \centering
    \includegraphics[width=0.48\textwidth]{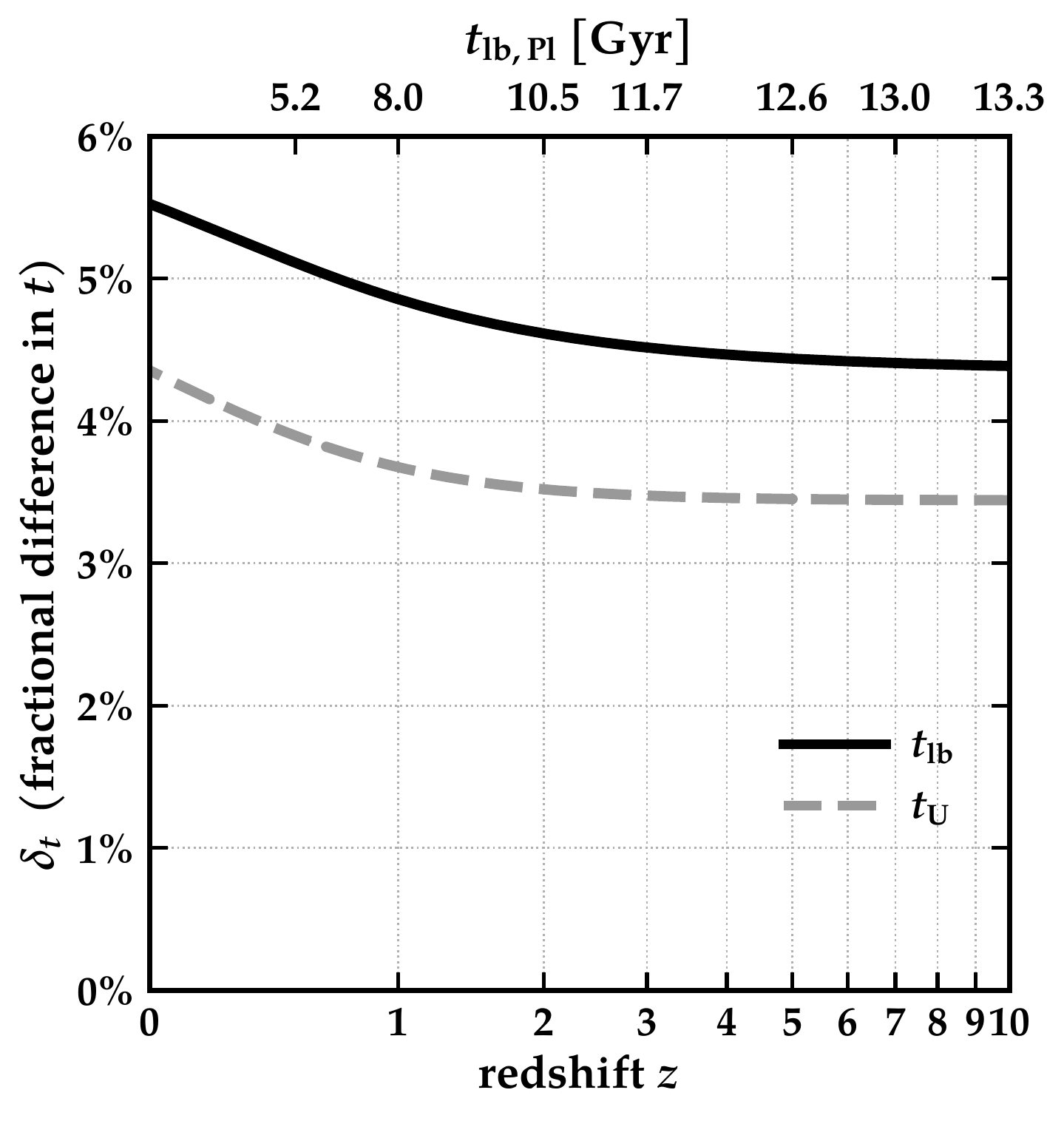}
    \includegraphics[width=0.48\textwidth]{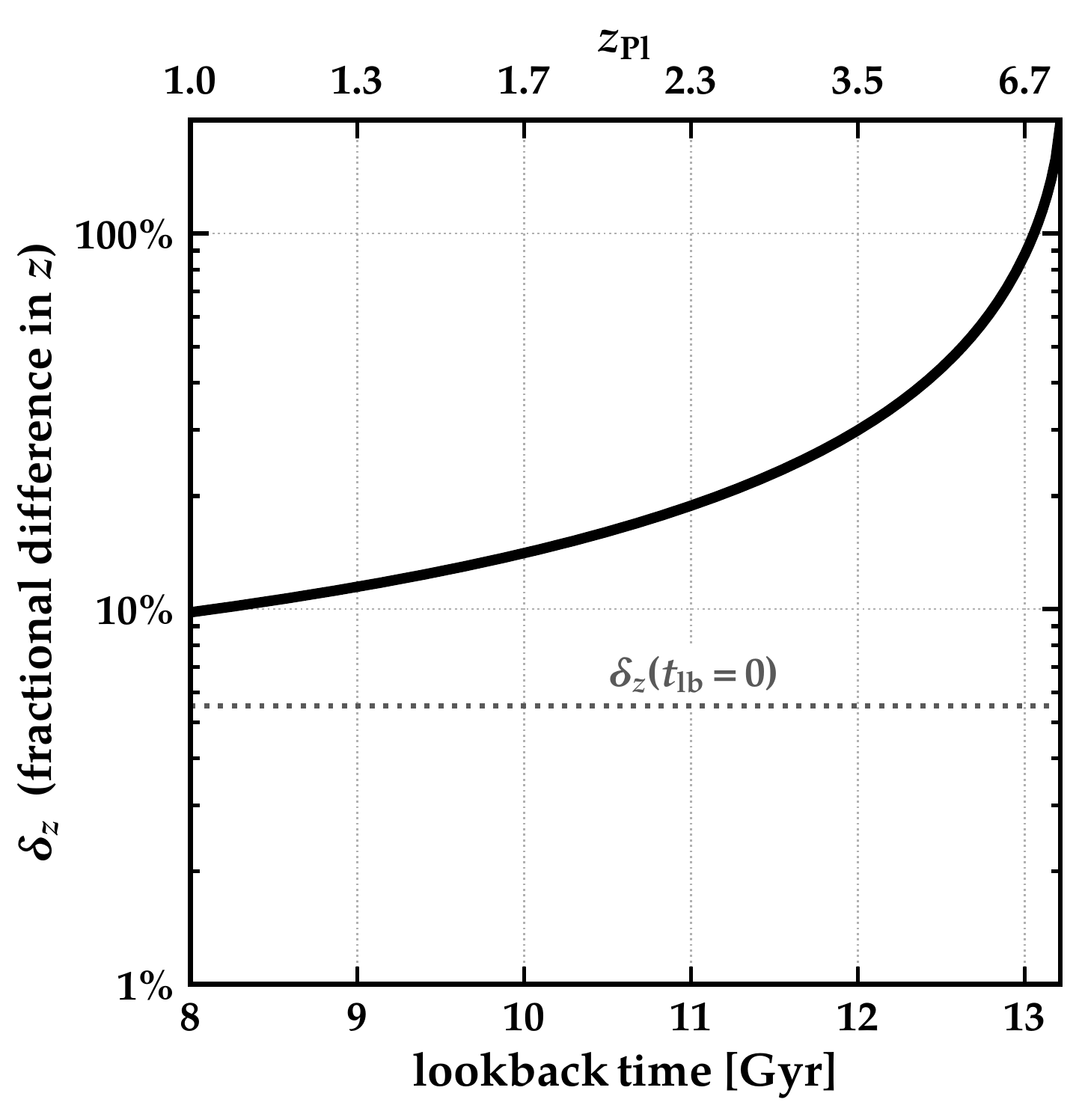}    
 \caption{\textit{Left:} The \textit{fractional} difference $\delta_t(z)$ in time as a function of redshift in EDE relative to \planck. The fractional difference in lookback time (black) and cosmic time (gray) between the two models is lowest at high redshift and increases at low redshift. The fractional difference in cosmic time always exceeds 3.4\% and reaches 4.4\% at $z=0$, while the fractional difference in lookback time is always larger than 4.4\% and reaches 5.5\% at $z=0$. 
 \textit{Right:} The fractional difference in redshift as a function of lookback time, $\delta_z(\tlb)$, over the range $8\, \gyr \leq \tlb \leq t_{\rm 0, EDE}$. The minimum value of $\delta_{z}(\tlb)$ is 5.5\% at low redshift, and it rises dramatically toward large values of $\tlb$, reaching 10\% at $8\,\gyr$ and $100\%$ at $13\,\gyr$. The very large redshift ranges of many entries in Table~\ref{tab:starages} have their origin in the large values of $\delta_z(\tlb)$ for $\tlb \gtrsim 12-12.5\,\gyr$.  \textbf{The redshift-time relationship has a current uncertainty of \textit{at least} $\bm{{\sim}4\%}$ at all $z$ and $t$.} 
 \label{fig:frac_diff_t}
}
\end{figure*}
%%%%%%%%%%%%%%%%%%%%%%%%%%%%%%%%%%%%%%%%%%%%%%%%%%%%%%%%%%%%%% 

The relationship between scale factor (or redshift) and cosmic time for a flat baseline model is given in Eq.~\ref{eq:t_of_a}, with Eq~\ref{eq:tlb_a} giving  corresponding relationship for lookback time. For $z \ll z_{\star}$, these relationships also hold at the sub-percent level for EDE models. Since Eq.~\ref{eq:tuniv_cosmo} depends on $(\om,\,h)$, the redshift-age relation differs in the two models. This is generally appreciated in the context of $\tuniv$, but it is important to note that ($\om, h$) affects the $z-t$ (or $z-\tlb$) connection affected at all redshifts and times.

Figure~\ref{fig:z_vs_z} compares the redshift that corresponds to a given lookback time in the \planck~cosmology ($x$-axis) and the EDE cosmology ($y$-axis). The color of the line shows the lookback time at each redshift, and selected fixed lookback times are shown on the plot to enable direct comparison. For $0 < z \lesssim 1$, the lookback time corresponding to a given redshift is similar in each model. As the lookback time gets larger, however, a systematic difference emerges, with fixed lookback time corresponding to a lower redshift in the Planck cosmology relative to EDE (because $t_{\rm 0,Pl} > t_{\rm 0,EDE}$): for example, a lookback time of 12.5 Gyr falls at $z \approx 4.6$ for Planck and $z\approx 7.2$ for EDE. The difference increases quickly, and dramatically, at even larger lookback times, with $z_{\rm Pl}=6$ giving the same lookback time --- $t_{\rm lb} =12.87\,\gyr$ --- as $z_{\rm EDE}=12.4$. 

An alternate way to look at the difference in the $z-t$ relationship in the two cosmologies is to plot the time difference as a function of redshift. Figure~\ref{fig:tlb_tuniv} shows $\Delta\,t_{\rm lb}$ (loobkack time; black solid curve) and $\Delta\,t$ (cosmic time; gray dashed curve) as a function of redshift, while the top $y$-axis give the lookback time in the \planck~cosmology corresponding to the redshift on the main $x$-axis. The difference in lookback times increases with increasing redshift, reaching $\Delta t_{\rm lb} \approx 0.4\,\gyr$ by $z=1$ (and asymptoting to $\Delta t_{\rm lb}=t_{\rm 0,Pl}-t_{\rm 0,EDE}$ as $z \rightarrow \infty$). The difference in cosmic time decreases with increasing redshift and is $\Delta t \approx 0.2\,\gyr$ at $z=1$; it also asymptotes to $\Delta t_{\rm lb}=t_{\rm 0,Pl}-t_{\rm 0,EDE}$, at $z=0$. 

As a complement to the absolute time difference, it is also useful to consider the fractional difference in time at a given redshift, $\delta_t(z)$, which we define as  
\begin{equation}
    \label{eq:delta_t}
    \delta_t(z) \equiv \frac{t_{\rm Pl}(z)-t_{\rm EDE}(z)}{[t_{\rm Pl}(z)+t_{\rm EDE}(z)]/2}
\end{equation}
for either lookback time or cosmic time. This is the quantity that can be used in assessing the accuracy of time measurements at a given redshift. The left panel of Fig.~\ref{fig:frac_diff_t} shows $\delta_t$ for both cosmic time (gray) and lookback time (black) as a function of redshift, with the upper $x$-axis giving $t_{\rm lb, Pl}(z)$. 
Unlike the absolute time difference, the fractional difference in times at fixed $z$ increases monotonically toward low redshift in both models, reaching 5.5\% for $t_{\rm lb}$ and 4.4\% for $t$. Importantly, the fractional difference has a minimum, non-zero value in both cases: at all redshifts,\footnote{We note that at $z \gtrsim 0.1\,z_{\star}$, Eqs.~\ref{eq:t_of_a} and \ref{eq:tlb_a} no longer hold in EDE, and a full calculation using the appropriate $H(a)$ (see Appendix~\ref{sec:appendix}) is required. Statements in this section should be taken to apply to $z \lesssim 0.1\,z_{\star} \approx 100$, which covers all directly measured astrophysical redshifts.} $\delta_t > 4.4\%$ for lookback time and  $\delta_t>3.4\% $ for cosmic time. 
These limits can be derived directly in terms of $\omega_{\rm m}$ and $H_0$ based on Eq.~\ref{eq:t_early} and \ref{eq:t_late}: $\delta_t=\delta(\tuniv)$ as $z \rightarrow \infty$ (for $\tlb$) and $z \rightarrow 0$ (for $t$), while $\delta_t=\delta(1/H_0)$ as $z \rightarrow 0$ (for $\tlb$) and $\delta_t=\delta(1/\sqrt{\omega_{\rm m}})$ as $z \rightarrow \infty$ (for $t$). The fractional difference in redshift at fixed lookback time, $\delta_z(\tlb)$ is shown in the right panel of Fig.~\ref{fig:frac_diff_t}. It has a minimum value of $\delta(1/H_0)$ ($\ssim 5.5\%$ for the models considered here) as $\tlb \rightarrow 0$. By a lookback time of $8\,\gyr$, the fractional difference in redshift between the two models is 10\%, and $\delta_{z}(\tlb=13\,\gyr)=100\%$. As we describe in \S~\ref{sec:examples} and Table~\ref{tab:starages}, the large difference in redshift at $\tlb \gtrsim 12\,\gyr$ has implications for using stellar ages to place constraints on cosmology or to reliably situate objects in specific cosmological epochs (e.g., the reionization era). 

The take-away from this subsection generally and Figure~\ref{fig:frac_diff_t} specifically is that the uncertainty in converting from a known redshift to time is at least 4.5\% for lookback time and 3.5\% for cosmic time. Any times obtained from converting from a known redshift to $\tlb$ or $t$ --- including the age of the Universe --- that are quoted to a higher level of precision do not reflect the current uncertainty in our understanding of cosmology. The uncertainty is actually somewhat larger, at least 5.5\%, when going from $\tlb \rightarrow z$. The redshift-time relationship, and our knowledge of the age of the Universe, will have an irreducible uncertainty of at least ${\approx} 4\%$ owing to uncertainties in the underlying cosmological model so long as 
effects such as those in the EDE example studied here 
cannot be ruled out by observations.

\section{Examples}
\label{sec:examples}
In this section, we consider examples of recently published stellar ages as directly reported in the original sources in order to illustrate both the promise and challenges of employing stellar ages as constraints on cosmology and galaxy formation.
We then discuss caveats, systematics, obstacles, and opportunities with various approaches to stellar age determination, a field that is the subject of a vast body of literature \citep[e.g.,][]{vandenberg1996, gallart2005, sneden2008, soderblom2010, cassisi2016, catelan2018}.  

The purpose of the discussion here is not to provide a comprehensive review of these topics, but rather to illustrate the potential, challenges, and confusion that the current generation of stellar ages poses, particularly in light of recent advances in cosmology. In \S~\ref{sec:discussion}, we provide a broader discussion of precision versus accuracy in stellar ages, prospects for improvement, and how stellar ages interface with current tensions between cosmological models.
In some cases, taking reported age error bars at face value results in seemingly implausible redshifts given any viable cosmological model and tensions with the cosmological age of the Universe. However, it is important to recall that the physics that determines a star's age 
is unrelated to the framework of cosmological models. 
This independence provides strong motivation for improvements in stellar age precision and accuracy, as well as scrupulous reporting of both.

\subsection{The reionization era}
%%%%%%%%%%%%%%%%%%%%%%%%%%%%%%%%%%%%%%%%%%%%%%%%%%%%%%%%%%%%%%
\begin{figure*}
 \centering
 \includegraphics[width=0.48\textwidth]{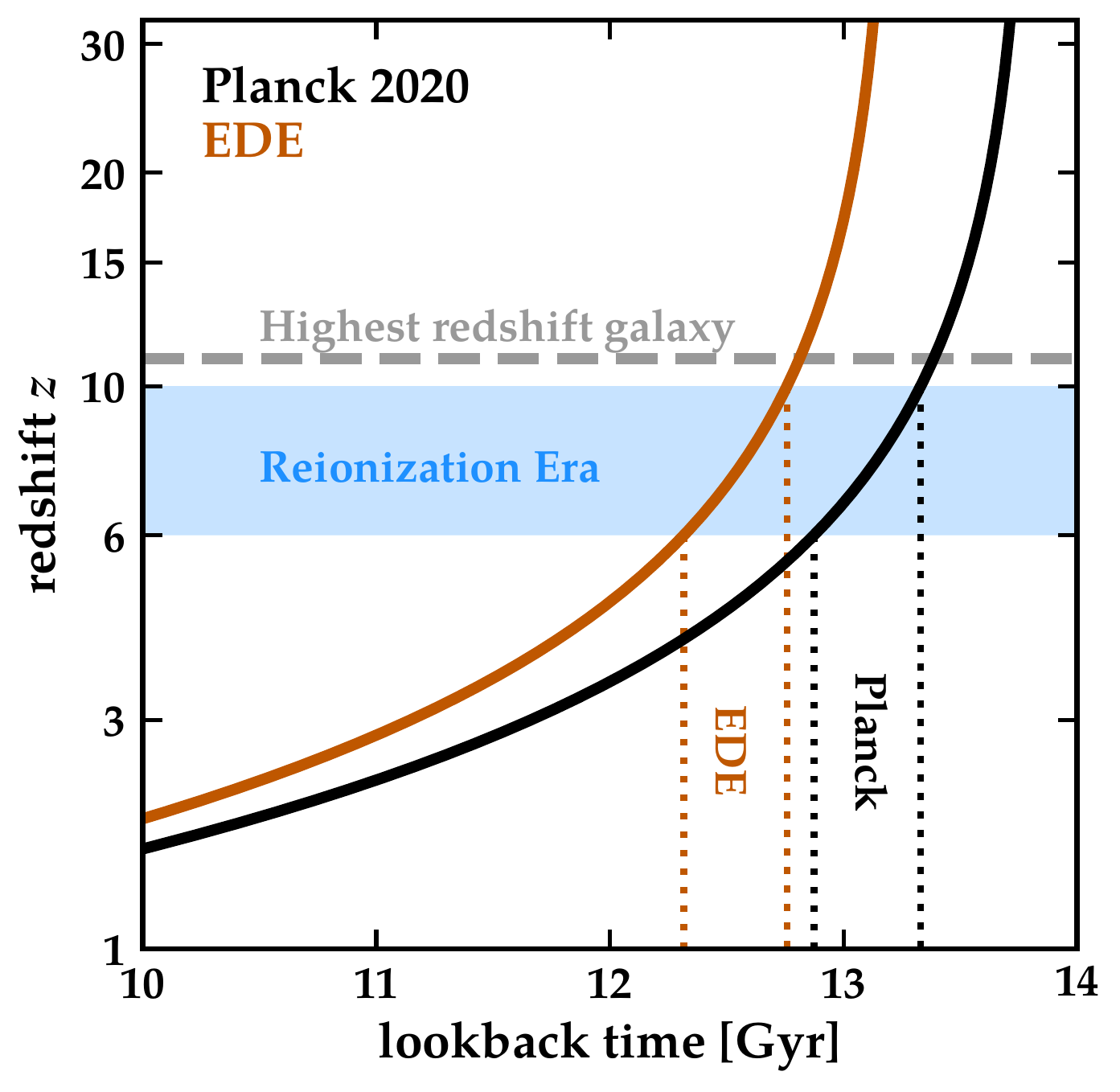}
 \includegraphics[width=0.48\textwidth]{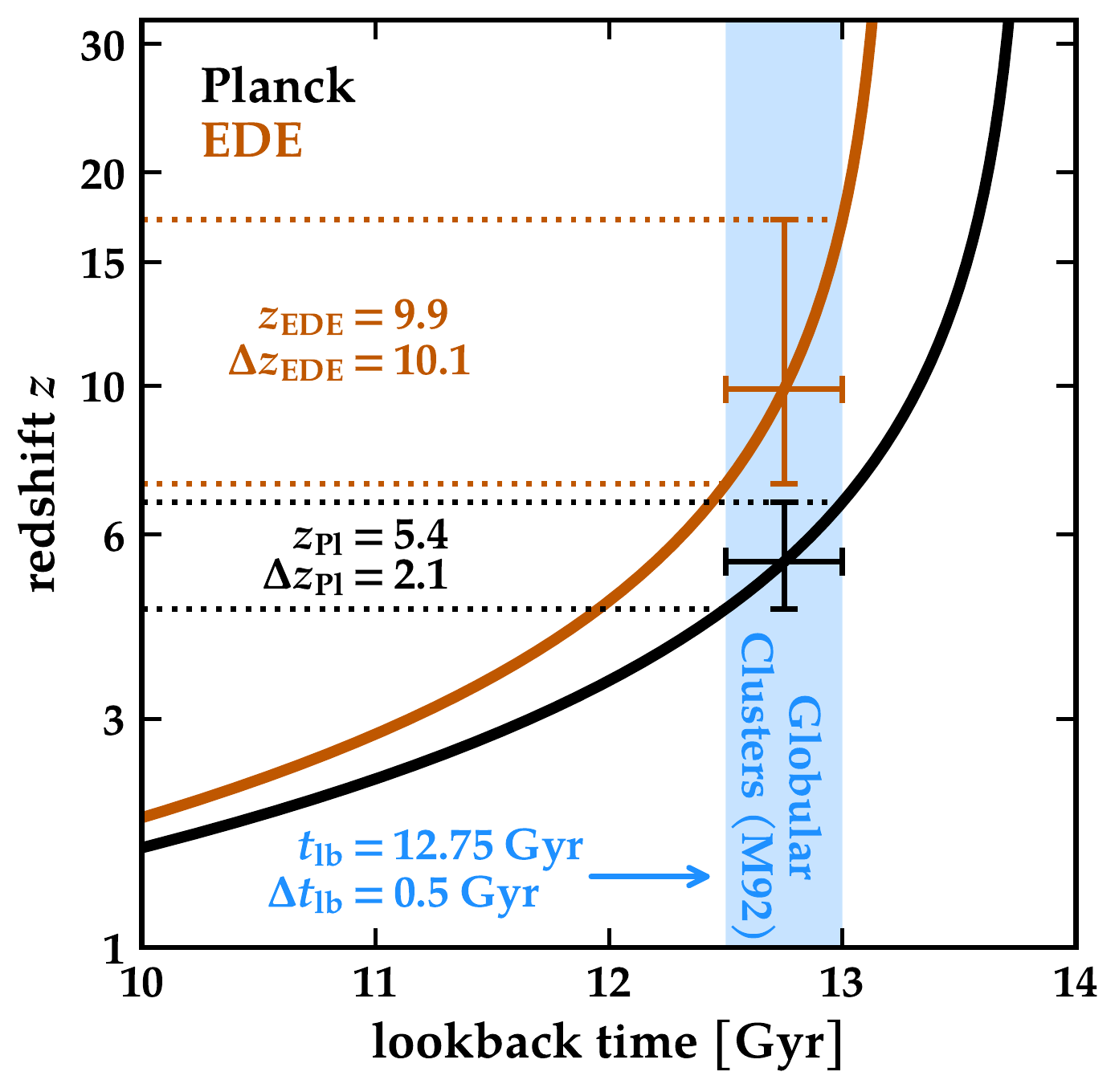}
 \caption{The relationship between lookback time ($x$-axis) and redshift ($y$-axis) for the \planck~(black curve) and EDE (dark orange curve) cosmologies. The left panel shows lookback times corresponding to a fixed redshift range, while the right panel shows redshifts corresponding to a fixed lookback time interval. \textit{Left}: the times in each cosmology corresponding to the reionization epoch ($10 > z > 6$). The reionization era has essentially the same duration in the two cosmologies but range of lookback times that reionization spans is disjoint in the two models, $13.33 > \tlb/\gyr  > 12.79 $ (\planck) versus $12.76 > \tlb/\gyr > 12.32$ (EDE). \textit{Right}: the redshifts in each cosmology corresponding to the formation time of a typical globular cluster, taken to be $12.7 \pm 0.25\,\gyr$ (which is a \textit{very} optimistic uncertainty range). This formation epoch corresponds to 
 $z_{\rm Pl}=5.44^{+1.28}_{-0.86}$ 
 versus $z_{\rm EDE}=9.87^{+7.34}_{-2.72}$, 
 i.e., it is the difference between globular clusters forming at the tail end of, or after, the reionization epoch (for \planck) and in the early phases of, or even previous to, reionization (for EDE).  
 \label{fig:t_vs_z}
}
\end{figure*}
%%%%%%%%%%%%%%%%%%%%%%%%%%%%%%%%%%%%%%%%%%%%%%%%%%%%%%%%%%%%%% 
We will use the epoch of reionization --- the period in the early Universe when the neutral fraction of the intergalactic medium transitioned from unity (i.e., fully neutral) to $<10^{-3}$ (i.e., almost fully ionized), roughly corresponding to $10 \gtrsim z \gtrsim 6$ (e.g., \citealt{stark2016, madau2017, greig2017}) --- as one important reference point for comparing ages and redshifts. 
Figure~\ref{fig:t_vs_z} illustrates how the timing of reionization in terms of lookback time depends on the adopted cosmology. In the \planck~cosmology, it took place from $12.87 - 13.33\,\gyr$ ago; in the EDE cosmology, the same redshift range corresponds to $12.32 - 12.76\,\gyr$ ago. 

One takeaway of Figure~\ref{fig:t_vs_z} is that an object (e.g., a star or globular cluster) with a well-defined and precisely known age cannot generically be a ``reionization-era'' object.  That is, because the EDE and \planck~cosmologies are equally well-fit by available cosmological data and have disjoint lookback times for the reionization era (Figure~\ref{fig:t_vs_z}), it is not possible to conclude with certainty that an object's age places its time of formation within the reionization era. 

Figure~\ref{fig:t_vs_z} also illustrates the effect that varying the cosmological expansion history has on determining ages for objects (e.g., high-redshift galaxies) with well-measured redshifts. For example, an object with a precisely known redshift in the middle of the reionization era has an uncertainty of $\ssim 550\,{\rm Myr}$ in lookback time. 

To explore this point further, consider the reported detection of emission lines corresponding to redshift $z=10.957 \pm 0.001$ in GN-z11 \citep{jiang2021}, a source originally detected photometrically with the \textit{Hubble Space Telescope} (\textit{HST}). This would make GN-z11 the highest redshift galaxy detected to date, and it is a pre-reionization-era object according to the definition adopted here. 
The redshift of GN-z11 (plotted as a gray line in the left panel of Fig.~\ref{fig:t_vs_z}) corresponds to $\tlb=13.39\,\gyr$ (cosmic time of $416$~Myr, and a radial comoving distance of $d=9.83\,{\rm Gpc}$) in the \planck~cosmology as opposed to $\tlb=12.81\,\gyr$ (cosmic time of $402$~Myr, radial comoving distance of $d=9.44\,{\rm Gpc}$) in the EDE cosmology. The same $\ssim4\%$ uncertainty in lookback time is also present in the distance of high-redshift objects.

\begin{table*}
% %%%%%%%%%%%%%%%%%%%%%%%%%%%%%%%%%%%%%%%%%%%%%%%%%%%%%%%%%%%%%%
\caption{Precision Ages for Select Near-Field Objects.  (1), (2) Name and type of the object. (3) Age of the object with uncertainties reported in the literature by (6); typically, these uncertainties reflect only the precision, and absolute uncertainties (i.e., accuracy) are either roughly estimated or not given (see \S~\ref{sec:examples}). (4), (5) The corresponding redshifts in the \planck\ and EDE cosmologies.  In some cases, the reported ages and redshifts may appear unrealistically high or the associated uncertainties may appear unreasonably small. We have elected to take the reported ages and uncertainties at face value and discuss the caveats in \S~\ref{sec:examples} and \S~\ref{sec:discussion}.}
\label{tab:starages}
\renewcommand{\arraystretch}{1.5}
\begin{tabular}{lccccc}
  \hline
  Object & Type & Age (Gyr) & $z_{\rm Pl}$ &  $z_{\rm EDE}$ & Ref \\ 
(1) & (2) & (3) & (4) &  (5) & (6) \\\hline
Sun & star & $4.567\pm 0.0016$ & $0.4164 \pm 0.00074$ & $0.4537\pm 0.00082$ & \citet{connelly2012} \\
\hline
J1312-4728 & star & $13.53\pm0.002$ & $14.88_{-0.08}^{+0.08}$ & --- & \citet{schlaufman2018} \\ %Dartmouth
& & $12.747\pm0.553$ & $5.43_{-1.58}^{+4.13}$ & $9.83_{-4.40}^{+\infty}$ & \citet{schlaufman2018}  \\ %Param1.3

& & $11.12\pm0.07$ & $2.44_{-0.06}^{+0.06}$ & $2.97_{-0.09}^{+0.09}$ & \citet{schlaufman2018}  \\ %MIST

CS 29497-004 & star & $16.5\pm6.6$ & ${>}1.66$ & ${>}1.90$ & \citet{hill2017} \\
 & & $13.7\pm4.4$ & $29.51_{-28.10}^{+\infty}$ & ${>}1.59$ & \citet{hill2017} \\
 RAVE J203843.2-002333
 & star & $13.0\pm1.1$ & $6.72_{-3.39}^{+\infty}$ & $17.22_{-12.79}^{+\infty}$ & \citet{placco2017} \\
 \hline 
 
 WD 0346$+$246 & WD & $11.49\pm1.51$ & $2.80_{-1.10}^{+3.92}$ & $3.52_{-1.57}^{+13.70}$ & \citet{kilic2012} \\

J1312-4728 & WD & $12.41\pm0.22$ & $4.34_{-0.50}^{+0.65}$ & $6.54_{-1.12}^{+1.79}$ & \citet{torres2021} \\

$8945908078561782540$ & WD & $13.949\pm0.845$ & ${>}7.47$& ${>}27.38$& \citet{fouesneau2019} \\

\hline 
M92 & GC & $12.75\pm0.25$ & $5.44_{-0.86}^{+1.28}$ & $9.87_{-2.72}^{+7.34}$ & \citet{vandenberg2013} \\
&  & $13.06\pm0.18$ & $7.13_{-1.10}^{+1.66}$ & $21.68_{-9.14}^{+\infty}$  & \citet{marin-franch2009} \\
&  & $13.25\pm1.0$ & $8.91_{-4.94}^{+\infty}$ & ${>}5.67$ & \citet{dotter2010}\\
& & $13.2$ & $8.35$ & $110.86$ & \citet{brown2014}\\

\hline 
Bootes~{\sc I} & UFD & $13.3\pm0.3$ & $9.56_{-2.83}^{+8.81}$ & ${>}17.22$ & \citet{brown2014} \\

CVn~{\sc II} & UFD & $13.6\pm0.3$ & $18.36_{-8.81}^{+\infty}$ & --- & \citet{brown2014} \\

Coma~Ber & UFD & $13.9\pm0.3$ & ${>}18.36$ & --- & \citet{brown2014} \\

Hercules & UFD & $13.1\pm0.3$ & $7.44_{-1.78}^{+3.80}$ & $26.72_{-15.98}^{+\infty}$ & \citet{brown2014} \\

Leo~{\sc IV} & UFD & $13.1\pm0.4$ & $7.44_{-2.19}^{+6.37}$ & $26.72_{-17.56}^{+\infty}$ & \citet{brown2014} \\

Ursa~Major~{\sc I} & UFD & $12.7\pm0.3$ & $5.25_{-0.93}^{+1.48}$ & $9.16_{-2.68}^{+8.06}$ & \citet{brown2014}
\\
\hline
 \end{tabular}
\end{table*}
% %%%%%%%%%%%%%%%%%%%%%%%%%%%%%%%%%%%%%%%%%%%%%%%%%%%%%%%%%%%%%%

\subsection{The ages of ultra-faint galaxies}
In galaxy formation theory, reionization plays the crucial role of setting the low-mass threshold of galaxy formation. The ionizing UV background heats the intergalactic medium to $(1-2)\times 10^{4}\,{\rm K}$; this is sufficient to prevent gas accretion onto halos below $M_{\rm vir}(z\ssim 8)=10^{8}\,\msun$, curtailing the supply of cold gas and inhibiting the formation of stars (e.g., \citealt{babul1992, efstathiou1992, thoul1996, gnedin2000, hoeft2006, okamoto2008, mcquinn2016, onorbe2017}). Observations of nearby ultra-faint dwarf (UFD) galaxies show strong evidence for nearly exclusively ancient stellar populations \citep{brown2014, weisz2014a, weisz2014b}, supporting this theoretical picture of a reionization-induced floor of galaxy formation that also helps reconcile the ``missing satellites problem'' \citep{klypin1999, moore1999} in \lcdm\ \citep{bullock2000, benson2002, somerville2002, ricotti2005, bovill2009, dooley2017, rodriguez-wimberly2019, wheeler2019}. 

The deep \textit{HST}-based color-magnitude diagrams of 6 UFDs acquired and analyzed by \citet{brown2014} are among the best evidence that UFDs are fossils of the reionization era.  The star formation histories of these systems indicate that all 6 systems formed the majority of their stars prior to reionization and stopped forming stars within $\ssim 1~\gyr$ of each other after reionization ended. Table \ref{tab:starages} lists the mean ages of these systems as listed in \citet{brown2014}. The reported errors on the mean ages of these UFDs reflect the $1\,\sigma$ uncertainties measured from isochrone fitting and suggest that the mean age of UFDs can be measured to a few percent precision. As an empirical check, \citet{brown2014} show that these UFDs appear to be as old as metal-poor Galactic globular cluster M92, for which they report an age of 13.2 Gyr. However, they note that in addition to their formal uncertainties, the ages of M92 and the UFDs may be uncertain in absolute age (i.e., accuracy) by up to $\ssim 1~\gyr$ owing to uncertainties in quantities such as distances, reddening, and stellar chemical abundance patterns.

For our purposes, the mean ages of UFDs serve two important purposes.  First, they are a clear point of comparison between reionization and stellar ages, if current \lcdm~galaxy formation theory is correct (i.e., if reionization quenches very low-mass galaxies).  
Second, they also provide a lower limit on the age of the Universe. In the \planck\ cosmology, the ages of all 6 UFDs are consistent with a formation epoch that is no later than the reionization era, and 
5 of the 6 galaxies are consistent with having formed before reionization. 
Ursa~Major~{\sc I} stands out as the sole exception: it has a mean formation redshift indicating that it formed during, or even slightly after, reionization. In comparison, UFDs formed at systematically higher redshift in the EDE cosmology, with all but Ursa~Major~{\sc I} consistent with being pre-reionization fossils. 

Perhaps the most striking characteristic for UFDs in Table \ref{tab:starages} is that 
several have mean ages that are uncomfortably close to, or greater than, the age of the Universe.
The best-fitting age of Coma~Berenices is $0.1~\gyr$ older than $t_{0, \rm Pl} = 13.8~\gyr$, and the best-fitting age of CVn~{\sc II} is a mere $0.2~\gyr$ younger than $t_{0, \rm Pl}$.
This tension is more pronounced in the EDE cosmology with $t_{0, \rm EDE} = 13.2~\gyr$: CVn~{\sc II} and Coma~Berenices are formally inconsistent with the age of an EDE Universe even when considering the quoted uncertainties.

One obvious solution is to include the additional $\ssim1~\gyr$ ($\ssim 7\%$) error suggested by \citet{brown2014} to account for uncertainties in quantities such as distance, reddening, and stellar abundance patterns.  Taking the extreme limit of this error --- shifting all mean ages younger by 1~Gyr --- places all UFD mean formation epochs within the age of the Universe for both cosmological models. However, such a shift complicates the interpretation of the expected connection between UFDs and reionization.  For example, if all mean ages of UFDs are shifted to be $\ssim 1~\gyr$ younger than listed in Table \ref{tab:starages}, then only Coma~Ber and CVn~{\sc II} are consistent with forming during or before reionization within the \planck\ cosmology, while the remaining 4 systems are all post-reionization fossils. Applying a similar shift to the mean ages in the EDE cosmology results in somewhat better agreement with expectations from galaxy formation theory, as only Ursa~Major~{\sc I} is inconsistent with forming at $z \gtrsim 6$.

In reality, shifting all ages by a uniform value of $\ssim 1~\gyr$ is an over-simplification and represents the extreme case.  The amplitude of the systematic uncertainties likely varies from object to object (e.g., as knowledge of their distances may be different).  The takeaway from this exercise is that current data on UFDs may capture the link between galaxy formation and reionization, but the current observational basis for this link is closer to suggestive than iron-clad. Shoring up the observational case for UFDs as fossils of the reionization era will require an investment in quantifying and reducing systematic uncertainties in age determinations. We discuss areas for improvement in these uncertainties in the context of globular cluster and stellar ages later in this section.

\subsection{The ages of globular clusters}
\label{subsec:globulars}
Unlike UFDs, which have broad metallicity distributions and extended star formation histories \citep{simon2019}, globular clusters (GCs) are thought to have stellar populations that are essentially single age and show little spread in [Fe/H] \citep[e.g.,][]{nardiello2015}. In theory, the age of a GC is well-defined, and a precise measurement of such an age can be used to set a lower limit on the age of the Universe and to help understand when, cosmologically, GCs formed. The cosmological utility of GCs has long been appreciated, first in terms of the historical $H_0$ debate (e.g., \citealt{tayler1986, vandenberg1996, chaboyer1996, chaboyer1996a}) and more recently 
in the context of GCs as tracers of galaxy and structure formation \citep{carlberg2002, renzini2017, boylan-kolchin2017, forbes2018, pfeffer2018, el-badry2019, adamo2020} and as potential contributors of ionizing photons during the reionization-era \citep{ricotti2002, schaerer2011, katz2014, boylan-kolchin2018}. 

The MW contains many GCs that appear to be $12-14~\gyr$ old (e.g., \citealt{vandenberg1996}). In the interest of brevity, we use a single and well-studied example, M92, to illustrate the current state of GC age determinations.  Table~\ref{tab:starages} lists ages from several widely-cited papers on GC ages \citep{marin-franch2009, dotter2010, vandenberg2013}, as well the age of M92 used to benchmark UFD ages \citep{brown2014}.  In general, the ages agree quite well, with a value of $13.00\pm0.25~\gyr$ ($\ssim 2\%$ precision) encapsulating the range of ages from these publications.  
Taking the reported ages at age at face value, M92 is consistent with being a relic of the reionization-era and forms comfortably within the age of the Universe in the \planck\ cosmology.  In an EDE framework, M92 is a pre-reionization relic and forms uncomfortably close to the beginning of the Universe ($\ssim200$~Myr after the Big Bang, on average). We illustrate this point in the right panel of Figure~\ref{fig:t_vs_z} with an M92-like cluster with an age of $12.75\pm0.25~\gyr$ (the \citealt{vandenberg2013} age and associated error for M92, and very close to the mean age of $12.7\pm 1\,\gyr$ for 9 old, metal-poor GCs found in \citealt{chaboyer2017}).

The above discussion of M92 only considers \textit{precision} in age. Similarly, the authors of all GC papers we consider in Table~\ref{tab:starages} caution that their ages and errors do not include the much harder-to-measure uncertainty in \textit{accuracy}. The most comprehensive study of age accuracy comes from a series of pioneering papers in which uncertainties in distance and extinction are formally considered in addition to our uncertain knowledge of stellar physics such as convection, opacities, and nuclear reaction rates \citep{chaboyer1995}.

Continuing in this tradition,
\citet[hereafter, \citetalias{chaboyer2017}]{chaboyer2017} determine the age of M92 by first calibrating the physics of stellar models to a set of metal-poor stars in the MW that have geometric parallaxes measured to $\ssim 1\%$ from the \textit{HST} fine guidance sensor.  These models are then used to measure the distances, ages, and ages of select GCs including M92. \citetalias{chaboyer2017} report the age of M92 to be $13.2\pm1.1$~Gyr; this error bar includes formal fitting uncertainties (precision) as well as uncertainties in the distance, reddening, and $\ssim15$ parameters that describe stellar interiors (accuracy).  \citetalias{chaboyer2017} repeat this exercise for several metal-poor clusters and report total uncertainties (combined precision and accuracy) of $7-10\%$.  \citetalias{chaboyer2017} do not explicitly report error components (e.g., how much is due to nuclear reaction rates), but they do note that distance, reddening, chemical abundance, and convection (characterized by mixing length) are the main contributors to the age accuracy, followed by select nuclear reaction rates. 

The absolute age of M92, $13.2\pm1.1~\gyr$, is therefore not sufficiently well known to determine if it formed pre- or post-reionization or if it is in serious tension with $t_{\rm 0, EDE}$.  We discuss the issue of absolute ages, and associated areas for improvement, in \S \ref{sec:discussion}.

\subsection{The Sun}
\label{subsec:theSun}
The age of the Sun is determined by analyzing the decay of long-lived radioactive elements found in Solar System meteorites.  \citet{connelly2012} report the age of the Sun to be $4.567\pm0.001~\gyr$  based `Pb-Pb age dating', which is shorthand for the decay of  ${}^{238}{\rm U}$ and ${}^{235}{\rm U}$ into ${}^{207}{\rm Pb}$ and ${}^{206}{\rm Pb}$.  This corresponds to a formation redshift of $z_{\odot, {\rm Pl}}=0.4169 \pm 0.00074$ or $z_{\odot, {\rm EDE}}=0.4464\pm 0.00080\,\gyr$. This is the rare (singular?) example of an object for which we know the age to much higher accuracy than we know the redshift.\footnote{In this case, we mean the redshift of formation. The Sun's cosmological redshift is known to high accuracy and precision (e.g., Aristarchus ca.~250 BCE, unpublished).}  

As detailed in \citet{connelly2017}, the uncertainty in the age of the Sun depends on the fidelity of the `Pb-Pb' age dating technique, which is not agreed upon within terrestrial laboratories to the level of $\ssim0.005~\gyr$.  Beyond uncertainties in the age-dating itself, there is some timeline ambiguity in the formation of the meteorites versus when the Sun begin its life on the main sequence.  This uncertainty may be as large as $\ssim0.05~\gyr$ \citep[e.g.,][]{sackmann1993}.

\subsection{Ancient stars and stellar remnants in the Milky Way}
\label{subsec:old_stars}
\subsubsection{Isochrone Fitting of Individual Stars}
\label{subsubsec:isochrones}
A common approach to stellar age dating is comparison of observed data to stellar isochrones. As an example of how isochrone fitting is commonly used in the literature, we consider the analysis of \citet{schlaufman2018}, who report the age of an ultra metal-poor (J1312-4728; [Fe/H]$=-4.1$; \citealt{melendez2016}) MW star residing in a binary system to be $13.535\pm 0.002$~Gyr.  This age is determined from exquisite data (i.e., high precision $\ssim15$~band photometry), a strong prior on the distance and chemical composition, and a broad knowledge of the mass (or mass ratio) from time series spectroscopy.  Essentially, this data is about as good as it gets when trying to measure the age of a typical star in the MW.  \citet{schlaufman2018} fit for age, distance, extinction, and metallicity using the Dartmouth \citep{dotter2008} isochrones with $[\alpha/{\rm Fe}]=+0.4$ and report a best fit age of $13.53\pm0.002~\gyr$, where the error bars reflect the formal fitting uncertainty (0.01\% precision) marginalized over all free parameters. 

This result suggests that J1312-4728 is among the oldest objects in the Milky Way and sets a stringent limit on $\tuniv$.  In the \planck\ cosmology, J1312-4728 would have formed at $z_{\rm Pl} \cong 14.9$, clearly prior to the epoch of reionization.  In contrast, J1312-4728 does not fit within the framework of an EDE cosmology.  Its very precise age of $\cong 13.53~\gyr$ is in ${>}100\,\sigma$ tension with the best-fit age of an EDE Universe, $t_{\rm 0, EDE}=13.2~\gyr$. While the $1\,\sigma$ uncertainty on $t_{\rm 0, EDE}$ is just at the edge of consistency with $13.53\,\gyr$, we note that this set of EDE parameters does \textit{not} resolve the Hubble tension, as it would require $H_0 < 70\,\kms\,{\rm Mpc}^{-1}$.

As a way to gauge the sensitivity of their fit to choice in stellar model, \citet{schlaufman2018} also fit this star's SED using solar-scaled PARSEC and MIST isochrones \citep{ bressan2012, choi2016} and find ages of $12.747\pm0.553~\gyr$ and $11.12\pm0.07~\gyr$, respectively.  These findings place J1312-4728 well within the age of both a \planck\ and EDE Universe and suggest it may have formed post-reionization. 

More broadly, this case study illustrates a challenge that will be come increasing common in the era of precision data for stars.  There is a wealth of data on J1312-4728: high precision photometry, secure knowledge of the chemical abundance patterns, minimal extinction, good constraints on the distances, and suitable fitting technique.  Moreover, \citet{schlaufman2018} compared the fit qualities (e.g., by evaluating the Bayesian evidence) among fits to different stellar models and concluding that the Dartmouth fit was far superior, formally speaking.  And yet, given the variation in ages between the different model, it is challenging to know if the age derived from Dartmouth-only fit can be taken literally for cosmological purposes.

Several analyses for the ages of ancient stars \citep[e.g.,][]{bond2013, vandenberg2014, omalley2017, chaboyer2017} suggest that stellar ages derived by the very reasonable methods in \citet{schlaufman2018} cannot be taken at face value.  For instance, using an HST-based parallax, \citet{bond2013} and \citet{vandenberg2014} find the age of metal-poor subgiant  HD 140283 to be $14.27\pm0.38~\gyr$ (2.6\%) with an additional uncertainty in the absolute age of $\ssim0.8~\gyr$ (5.6\%) owing to bolometic corrections, abundances uncertainties, etc.  These papers do not, however, varying mixing length, which, as pointed out by \citet{vandenberg2014} and \citet{chaboyer2017} (among others), will affect the absolute age determination.   \citet{omalley2017} and \citetalias{chaboyer2017} conduct a more comprehensive fitting (i.e., including varying mixing length, nuclear reaction rates, etc.) of nearby metal-poor stars with \textit{HST}-based parallax measurements and find typical absolute uncertainties of $1-1.5~\gyr$ on the ages of individual stars ($7-10\%$).  Though computationally demanding, these types of studies provide a template for how to measure absolute stellar ages.  We discuss prospects for doing so on a larger scale in \S~\ref{sec:discussion}.

\subsubsection{Nucleocosmochronometers}
\label{subsubsec:cosmochron}
It is possible to measure ages for metal-poor stars that are enhanced in $r$-process elements by using abundances of radioactive elements with half-life decay times of several Gyr or longer (e.g., U, Th), a technique known as `nucleocosmochronology' that is long-established in the literature \citep{fowler1960, butcher1987, cowan1991, cowan1991a, cayrel2001, sneden2008}. Table \ref{tab:starages} lists two illustrative examples of stars with ages measured from radioactive decay.  

One example is CS 29497-004: using $R\ssim75000$ optical spectroscopy, \citet{hill2017} measured 46 elements in total and 31 r-process elements that include U and Th, which serve as age indicators when compared to one another and to more stable elements.  Using U/Th alone, \citet{hill2017} find an age of $16.5\pm6.6~\gyr$ for CS 29497-004, which translates into lower redshift bound of $z_{\rm Pl} > 1.66$ and $z_{\rm EDE} > 1.90$. The old age and large uncertainties are primarily driven by the low S/N of U absorption in the spectrum. \citet{hill2017} also report an age of $13.7\pm4.4\,\gyr$ for CS 29497-004, which is the result of averaging over several abundance ratios (i.e., Th/X and U/X).  
\citet{placco2017} provide another radioactive age determination of metal-poor star RAVE J2038-0023.  With $R\ssim66000$ optical spectroscopy, they measure 24 r-process elements, including U at $S/N > 100$ and report a U/Th age of $13.4~\gyr$ and a mean age of $13.0\pm1.1~\gyr$ from averaging over various abundance ratios (i.e., Th/X, U/X), which translates into redshifts of $z_{\rm Pl} = 6.72_{-3.39}^{+\infty}$ and $z_{\rm EDE} = 17.22_{-12.79}^{+\infty}$. 

The strength of nucleocosmochronolgy is that the radioactive decay times of isotopes, particularly for U, are extremely well-known and are largely invariant to underlying physics assumptions \citep[e.g.,][]{hill2002}.  However, U is extremely challenging to measure in all but a small fraction of stars, and even in those cases exceedingly good data is required (e.g., SNR$\gtrsim$100 and $R\gtrsim50000$).  

\subsubsection{Ancient White Dwarfs}
\label{subsubsec:WDs}
As the remnants of low-to-intermediate mass stars, white dwarfs (WDs) span a range of ages that can date back to the very early Universe. The relatively simple physics of WDs makes it possible to determine their cooling age to good accuracy and precision. When combined with knowledge of their progenitor's stellar mass, and hence lifetime, WDs have the potential to determine ages independent of cosmology and certain aspects of stellar physics (e.g., various quantities that set the luminosity of the main sequence turnoff; MSTO) that are relevant for some of the methods discussed earlier in this section \citep[e.g.,][]{schmidt1959, winget1987, fontaine2001}. 

The cooling of a WD depends on its mass (or radius), temperature, and atmospheric composition \citep{dantona1990}, and cooling ages are usually inferred by modeling the luminosity function and/or color-magnitude diagram, or through spectral energy distribution fitting \citep{hansen2003}. Historical limitations in WD age estimates include uncertainties in WD distances, and hence luminosities, and in incompleteness due to how faint they are, which meant that the faintest (and usually oldest) WDs could be missing.  Modern facilities and surveys (e.g., SDSS, \textit{HST}, Gaia) have largely mitigated these challenges, and now large populations of WDs are found in a number of GCs and in the field (e.g., \citealt{kleinman2013, gentile-fusillo2019}) enabling detailed analyses of their masses, compositions, and ultimately ages \citep[e.g.,][]{bergeron2019}.

Historically, ages of WDs have been measured in GCs, as they provided good estimates for distances and an independent comparison point for age. The progenitors of the WD cooling sequence in a given GC should all have the same age, meaning that if it is possible to detect the faintest WDs on the cooling sequence, it is possible to measure the age of the GC. \textit{HST} has enabled such detections (e.g., \citealt{hansen2007, bedin2009, hansen2013, campos2016}). The resulting ages derived from fitting the WD cooling sequence have precision of $\ssim 4-5\%$, but systematic errors are at least a factor of two larger owing to uncertainties in distances, reddening, and chemical composition \citep{campos2016}.

Outside of clusters, Table \ref{tab:starages} provides a few examples of ages for individual ancient WDs.  For must of the past decade, WD~$0346+246$ was one of the oldest known WDs in the Galactic disk.  \citet{kilic2012} find a total age of $11.49 \pm 1.51~\gyr$, which places its formation broadly during the cosmic noon epoch in both \planck\ and EDE cosmologies.  More recently, \citet{torres2021} report and age of $12.41 \pm 0.22~\gyr$ for J1321-4728, which is located in the Milky Way's halo.  In a \planck\ cosmology, it formed between reionization and cosmic noon, whereas in an EDE cosmology, it formed during the epoch of reionization (c.f.~Figure~\ref{fig:t_vs_z}).  

\citet{fouesneau2019} determined the ages of $\ssim100$ WDs in WD-main sequence wide binaries using a combination of Gaia DR1 parallaxes with optical and near-infrared photometry.  Several of their WDs have ages that are near or exceed the cosmological age of the Universe.  For example, they report that $8945908078561782540$ has a mean age of $13.939 \pm 0.845~\gyr$.  This age is within $1\,\sigma$ of the \planck\ age of the Universe but is uncomfortably close to $t_{\rm 0,EDE}$.  In both cases, this WD is a reionization-era relic (or even older), and by extension, its MS companion star would also be among the oldest stars in the Galaxy. 

\citet{fouesneau2019} emphasize that they only report \emph{relative} ages and
that their ages, in an absolute sense, are uncertain at the $5-10\%$ level.  In
the example above, an additional $5-10\%$  error translates to $\ssim 0.7-1.4~\gyr$ in lookback time and would make this WD (and its MS companion) younger than the age of the Universe, and possibly a post-reionization object, in both \planck\ and EDE cosmologies.  \citet{fouesneau2019} note that the predominant random error is parallax uncertainties.  Given the vast improvement in parallax provided by Gaia DR3 over DR1, it is possible that the precision on the age this WD can now be determined to $1-2\%$.  While they provide an estimate of the absolute age uncertainty, \citet{fouesneau2019} do not attempt to quantify it.

In general, WDs have vast potential as chronometers in a cosmological context.  Owing largely to Gaia, we now have large collections of WDs in the field with well-determined parallaxes and luminosities.  Because distance has been the main limitation to date, it will soon be routine to measure cooling ages of WDs to $\lesssim1\%$ precision. The accuracy of WD cooling ages will ultimately depend on the underlying physical models of WDs. This is promising, as WDs are in many ways much simpler systems than stars. Non-trivial uncertainties do remain, particularly with regards to how sedimentation, atmospheric properties, neutrino emission, and crystallization affect WD cooling rates \citep{van-horn1968, mochkovitch1983, segretain1994, bildsten2001, cheng2019, blouin2020}. However, the aforementioned Gaia observations are providing data with significant constraining power (e.g., \citealt{tremblay2019, bauer2020}), and some uncertainties may be mitigated by focusing on the most metal-poor clusters (e.g., crystallization uncertainties are larger for higher ${}^{22}{\rm Ne}$ abundance). Encouragingly, differences in the theoretical model predictions for WD cooling curves for a given set of physical assumptions appear to be at the few percent level \citep[e.g.,][]{salaris2013}. While uncertainties in the underlying physics are significantly larger, there is reason for cautious optimism that significant improvements are possible: for example, \citet{caplan2021} discuss updated diffusion coefficients applicable to WDs that are accurate to 1\%, a substantial advance relative to the 10\% uncertainty of previous calculations.

Other systematics include the requisite assumption of an initial mass -- final mass relationship \citep[e.g.,][]{kalirai2008, catelan2008, salaris2009, el-badry2018c}, which may be the single largest source of error, and knowledge of the main sequence lifetime of the progenitor star. The latter is somewhat sensitive to stellar physics, but not as much as quantities such as the luminosity of the main sequence turnoff. Simultaneously fitting the WD cooling sequence and the main sequence is possible in some GCs, which can provide provide a valuable cross-check on the two methods and might help to break degeneracies in models. In particular, while distance and reddening are still required parameters, they cannot differ for the WDs and main sequence stars in the same GC, which may provide tighter constraints on model fits.

\section{Discussion}
\label{sec:discussion}
A central theme to this paper is the age-old issue of precision versus accuracy.  Precision provides for constraints on the particulars of a given model, while accuracy is a statement about how well a given model approximates nature. A prime example of precision science is cosmology from \textit{Planck}, which provides better than percent-level precision on most base parameters of the \lcdm\ model.  However, the Hubble tension and its potential resolution via EDE indicate that \lcdm\ may not be an accurate (or at least not complete) description of reality.  Importantly, it is only because cosmology has become such a precise science that we can begin to ask and answer plausible questions about the accuracy of cosmological models.   We are poised to enter a comparable era of precision stellar astrophysics, which we believe will eventually lead to improvements in the accuracy of stellar theory and ages.  

\subsection{Towards an accurate cosmology}
It is tempting to draw an exact parallel between the current Hubble tension and the famed 50 versus 100 debate of the 20th century \citep{rowan-robinson1985}. While similarities exist, the fundamental difference is that \lcdm~is firmly entrenched as the default model at present and the Hubble tension directly challenges the model's completeness, whereas no such baseline model existed in, e.g., 1985. A precise determination of $h=0.5$ or $h=1.0$ in the 1980s would have pointed toward specific physics needed for establishing a standard cosmological model.
Having an established and well-tested model allows us to probe for cracks in that model, and small deviations from a theory's predictions are only meaningful if that theory is able to make precise predictions. Although cosmology is not at the level of the Standard Model of particle physics, where predictions are so precise that deviations at the level of 0.1 part per million can indicate the need for new physics, it is exactly because we have sub-percent-level measurements of various cosmological quantities that seemingly small deviations can be momentous.

Our current lack of understanding, at a fundamental rather than phenomenological level, of dark matter and dark energy leaves plenty of room for physics not encapsulated by the base \lcdm~model \citep{peebles2002}. And yet, those deviations are surprisingly strongly constrained in many cases (see, e.g., \citealt{knox2020, balkenhol2021}), with data providing broad consistency with base \lcdm~across a range of scales and redshifts. The unquestionable success of the base \lcdm\ model in explaining the large-scale ($k \lesssim 0.2\,\mpc^{-1}$) distribution of matter and energy in the Universe from the epoch of last scattering to the present day means that any ultimate cosmological model must look very similar to \lcdm~on those scales and at those redshifts. 

However ``very similar to'' is not ``the same as''. The Hubble tension is one possible indication of the incompleteness of cosmology's standard model, (base) \lcdm. And apparently very minor deviations from \lcdm\ --- in this case, an always-subdominant component of dark energy that is less than 1\% of the critical density for all of cosmological history except for 1 decade in scale factor, $1.5\times 10^4 \gtrsim z \gtrsim 2 \times 10^3$, can lead to the measurable effect of shifting the sound horizon by 4\%. An attendant effect, as we have discussed in detail, is a shift in the redshift-time relation of the same magnitude. Given this perhaps surprisingly large effect from a seemingly small change to the expansion history, other potential revisions to the base \lcdm~model are intriguing to consider. 

Perhaps the simplest example of a revision to the base \lcdm~model is to relax the assumption of flatness (adding $\Omega_{\rm k}$ as a parameter). In this extension, the same analysis of \textit{Planck} data as we consider for the base \lcdm~model results in (1) an overall fit that is not appreciably better than the base \lcdm~model; (2) a slight negative curvature, $\Omega_{\rm k}=-0.0106\,(-0.0092)\,\pm 0.0065$; and (3) a shift in the inferred value of the Hubble constant to $H_0=63.6\,(64.03)\,\pm 2.2\,\kms\,\mpc^{-1}$ \citep{planck2020}. The precision of this $H_0$ measurement is a factor of $\ssim4$ worse than in the base \lcdm~model, which underscores the model-dependence of precision. The coming era of large galaxy surveys and CMB experiments \citep{desi-collaboration2016, amendola2018, ivezic2019, wfirst2015, cmbs4_2019} will likely place us firmly in the realm of assessing the accuracy of the \lcdm~model and hopefully revealing the underlying physics of the components that are currently known only phenomenologically (or perhaps even overturning the entire \lcdm~paradigm).

\subsection{Towards accurate stellar ages}
In the context of stellar age measurements, the difference between precision (also known as relative ages) and accuracy (absolute ages) is a well-studied, decades-old issue.  The excellent review articles by \citet{stetson1996} and \citet{vandenberg1996} recount the storied history of relative and absolute ages of GCs, respectively. \citet{stetson1996} succinctly captures the essence of relative ages and why they are more commonly reported in the literature: ``Relative age determinations [can] use stellar evolution theory in a strictly differential sense, removing most of the effects of theoretical uncertainties in absolute chemical-abundance ratios, opacities, convection formalism, temperature-color relations, and the like. Differential comparisons can also be devised which reduce the effects of observational uncertainties in the absolute distance scale, overall metal abundance, and individual cluster reddenings.''  Relative ages can therefore address issues such as the formation chronology of MW GCs, whereas absolute ages are required to use GCs (or any stellar object) in a cosmological context (e.g., at what redshift did a GC form? or, how old is the Universe?). 

While relative ages establish a set of ages for differential comparisons given a fixed set of assumptions about stellar physics and observational uncertainties, changes in these underlying assumptions result in a new set of ages that could have different relative values. In contract, absolute ages and their associated uncertainties should naturally encompass allowable changes in the underlying parameters. For example, if the absolute age of a GC is $12\pm 1~\gyr$, any reasonable variations in the distance, reddening, and/or stellar physics should be captured by the stated uncertainty. Absolute ages are therefore more challenging to compute, have larger uncertainties, and are not always needed for a particular science goal; accordingly, relative ages are more commonly reported in the literature, though they are not always explicitly labeled as such.

The measured age of a star depends on knowledge of its distance, line-of-sight extinction, chemical composition, and the adoption of a stellar model (which, for simplicity, includes stellar interiors and atmospheres).  Many ages in the literature choose a fixed stellar model, and vary some permutation of the first three quantities.  We classify the results of this approach as a relative age, i.e., a measure of precision. This is because in many cases, distance/reddening/chemical composition affect the fundamental properties of a star (e.g., luminosity) in ways that are not entirely orthogonal to the underlying stellar physics.  For example, \citet{dotter2017} demonstrate that the inclusion of heavy element diffusion affects the shape, color, and temperature of the main sequence turnoff, and hence age, for stars of all ages.  Fitting data with stellar models that do not include diffusion result in stellar ages and/or metallicites that vary by up to 20\% compared to when diffusion is included.  Convection is another example.  Most stellar models tune their treatment of convection, usually via mixing length theory \citep[e.g.,][]{bohmvitense1958}, to the Solar value.  But several recent studies suggest that mixing length may vary with stellar parameters \citep[e.g., surface gravity, chemical composition;][]{trampedach2011, bonaca2012, tayar2017}, ultimately changing the temperature, luminosity, and/or size of a star and hence its inferred age and composition (\citetalias{chaboyer2017}, \citealt{valcin2021}).  

The \citet{schlaufman2018} example in \S\,\ref{sec:examples} illustrates the conundrum of modern stellar age determination and reporting. They have exquisite data, place a strong Gaia parallax prior on distance, and employ sound statistical techniques. Yet, they find three ages that are each very precise to $\la 4$\%, but are entirely disjoint.  Their (highly reasonable) solution is to report a best age estimate by comparing Bayesian evidence among the fits and asserting that only the Dartmouth models had an $\alpha$-enhancement that is similar to what is known from spectroscopy.  This process results in preferred age of $13.53\pm0.002~\gyr$, which is in tension with $t_{\rm 0,EDE} = 13.2~\gyr$.  One could go a step further and use the spread in ages among the three fits as a proxy for the absolute age uncertainty \citep[e.g.,][]{dolphin2012}, which would be $\ssim1.5~\gyr$ in this case.  But this also is not very satisfying, as the other models considered are clearly not as well-matched to the data, e.g., they use Solar-scaled rather than  $\alpha$-enhanced abundances, making this an overly conservative age uncertainty.  It is even more complicated when attempting to compare ages, and their uncertainties, across the literature due to the adoption of varying fitting techniques, model choices, what gets reported as an uncertainty, and even how an uncertainty is described (i.e., there is some confusion between what qualifies as an absolute age).

Fortunately, the stellar data revolution is making the path for improving relative and absolute ages very promising.  Beyond the wealth of well-calibrated multi-band stellar photometry and a plethora of spectroscopy, our knowledge of distances \citep[e.g.,][]{bailerjones2018, brown2018, chen2018, neeley2019, maizappellaniz2021, soltis2021}, extinctions \citep[e.g.,][]{schlafly2016, green2018, green2019}, and abundance patterns (see \citealt{jofre2019} and references therein) for large samples of stars, clusters, and remnants is improving such that relative ages can (will) be routinely measured to the percent/sub-percent level.

Much like how precision cosmology has enabled stress tests of \lcdm, the era of stellar precision will yield vast improvement in absolute stellar age determinations.  The approach pioneered by \citet{chaboyer1995}, in which stellar age/parameter determinations include varying the underlying stellar physics, provides guidance for moving forward. It is becoming computational tractable to generate large sets of stellar models that include variations in the (uncertain) underlying physics and to then fit these large sets of models to data.  This process becomes invaluable if it can be done in the context of sampling posterior distributions (e.g., via Markov chain Monte Carlo approaches), as it will not only yield absolute stellar ages, it will also provide quantitative constraints on --- and correlations among --- parameters of the underlying underlying stellar physics, conditioned on exquisite data.

Based on the results of \citetalias{chaboyer2017}, in which absolute age uncertainties are estimated to be $7-10\%$, it is likely that current and upcoming data, distances, and modeling techniques will soon enable measurement of the ages of stars, stellar remnants, and GCs to sub-percent level precision and $\ssim 4-5\%$ accuracy, which is comparable to the current levels of cosmological age accuracy.  With such good accuracy, stellar ages will be able to place important constraints on the history of our Universe at $z \la 15$ independent of an assumed cosmology.  Moving beyond this level of accuracy for stars will likely require a re-examination in the fundamental ingredients of current generations of stellar models \citep[e.g.,][]{arnett2015} and moving toward fully 3D simulations of MSTO and sub-giant stars.  The prospect of highly accurate ages is perhaps even more promising for WDs owing to simpler physics and a good level of agreement in ages among newer generations of models. A full analysis that considers variations in all parameters of current WD models, along the lines of \citetalias{chaboyer2017}'s work on MSTO-fitting for globular clusters, would be highly valuable in establishing the state-of-the-art in WD age accuracy.

\subsection{Interpretation}

The cosmological age of the Universe has long been an important benchmark when measuring the ages (and associated uncertainties) of ancient stars \citep[e.g.,][]{vandenberg1996}.  Taken at face value, cosmological age of the Universe is known to $\ssim0.2$\% from \textit{Planck} data, whereas stellar ages are known to $7-10\%$, a factor of $\ssim20$ worse. This suggests that the ages of ancient stars are relatively poorly known and that it would take dramatic --- and perhaps even impossible --- levels of improvement in our understanding of stellar physics and distances to stars for stellar ages to be competitive with cosmological ages.

The situation is more complicated, however. The $7-10\%$ uncertainty in stellar ages are a reflection of accuracy, whereas a $0.2$\% uncertainty in the \textit{Planck}-based age of the Universe is a measure of precision. Another point of comparison could be between the measures of precision for stellar and cosmological ages.  For example, compare the preferred age for J1312-4728 ($13.53\pm0.002~\gyr$; 0.01\%) from \citet{schlaufman2018} to either $t_{\rm 0, Pl} = 13.797\pm0.023~\gyr$ and $t_{\rm 0, EDE} = 13.246\pm0.17~\gyr$.  This indicates roughly comparable precision for ages of ancient stars --- in the presence of comprehensive data --- and the age of the Universe from cosmology. 

Yet, this is still an inapt comparison: changing a component of the underlying stellar model, e.g., the mixing length or chemical abundance patter, will change the age of J1312-4728 by more than 0.1\%, as is shown in \citet{schlaufman2018}, when different stellar models are used in the fitting. At this point, any comparison between precision stellar ages (or between stellar and cosmological ages) requires significant context.

The EDE solution to the Hubble tension illustrates one plausible path for 
comparing stellar and cosmological ages, including uncertainty estimates. 
EDE effectively introduces a set of additional parameters that go beyond the conventional base \lcdm\ model. In stellar modeling, one could modify the modeling of convection to vary the mixing length parameter (or introduce additional parameters to more fully capture the effects of convective mixing). This is not standard practice in stellar age-dating, but there is a common understanding in the literature that such parameters, especially convection and/or mixing length, do affect the determination of other parameters, including age \citep[e.g.,][]{chaboyer1995, vandenberg1996, vandenberg2013, kupka2017, chaboyer2017}.
Invoking EDE or other effects that affect the expansion history or interactions of matter and energy in the Universe has not been standard practice either (though cosmological analyses often search for indications of specific modifications to the base \lcdm~model), yet it affects parameter estimation at a level that is larger than the quoted precision. 

Perhaps the fairest age comparison, then, is between the accuracy of stellar ages 
when including variations in stellar evolution parameters and distance/reddening to the accuracy of cosmological ages when considering allowed extensions to the base \lcdm~model such as EDE or non-zero curvature. 
In this view, stellar ages can be constraining in the context of cosmological models if stellar age accuracy reaches the level of $4-5\%$, which, as we have argued in \S~\ref{sec:discussion}, appears plausible in the foreseeable future. 

\section{Conclusions}
\label{sec:conclusions}
Cosmological ages and distances are only defined in the context of a cosmological model that allows us to link the expansion history of the Universe to its energy constituents. In the context of the 6-parameter base \lcdm~cosmological model, \textit{Planck} data provide sub-percent precision on most cosmological parameters, including the $z \leftrightarrow t$ correspondence. However, the completeness (accuracy) of the base \lcdm~model is called into question by the Hubble tension. In this paper, we quantify the existing cosmological uncertainties in the redshift-time relation, using early dark energy as representative possible resolution of the Hubble tension, and explore the possibility of using stellar ages as an alternate constraint on cosmological models. Our main conclusions include:
\begin{itemize}
    \item The redshift-time relation is known to no better than $3.5\%$ for cosmic time and $4.5\%$ for lookback time (the relevant quantity for comparing to stellar ages) at all times and redshifts (Fig.~\ref{fig:frac_diff_t}).
    \item This uncertainty affects astrophysical interpretations of age determinations of nearby systems: for example, a star, GC, or UFD with a very precisely known age cannot generically be considered a reionization-era relic, as the reionization era spans disjoint lookback periods in the two cosmologies (Fig.~\ref{fig:t_vs_z}). This complicates galaxy formation theory interpretations of these objects and their relationship to reionization.
    \item The age of the Universe is very precisely determined in the base \lcdm~model because of a tight correlation between the angular size of the sound horizon at last scattering and $\tuniv$; EDE does not have this same tight correlation, and the resulting uncertainty in $\tuniv$ is a factor of 8 larger. 
    \item The best-fitting age of the Universe for the EDE model considered here is uncomfortably low relative to the reported ages of stars, globular clusters, and ultra-faint dwarf galaxies. High-age EDE models are in less tension with stellar ages, but these also require low $H_0$, meaning such models would not resolve the Hubble tension (Fig.~\ref{fig:t0_h0}).
    \item The existence of a $\ssim 5\%$ uncertainty in the cosmological $z \leftrightarrow t$ relation argues that measurements of stellar ages \emph{should not} adopt the cosmological age of the Universe as a strong prior for stellar models. 
    \item The era of large, well-calibrated stellar data (e.g., precise and accurate distances, photometry, abundances) and related modeling approaches have the potential to improve accuracy on stellar ages from the current $\ssim 7-10\%$ to $\ssim5$\%, making them competitive with cosmology.  
\end{itemize}

The central point emphasized here --- that the redshift-age relation from cosmology is uncertain at the $\ssim 4\%$ level --- is only possible to identify because of the level of precision of modern cosmological measurements. It can actually be seen as a remarkable success of the base \lcdm~model that EDE is so well constrained. Similarly, it is because we are now in the era of percent-level precision for stellar parameters that it is possible to closely examine the issue of accuracy of stellar ages. An exciting prospect is that in the coming era where large galaxy surveys and CMB experiments promise to provide even stronger cosmological constraints, accurate stellar models will serve as orthogonal tests of the physics of the Universe.

\section*{Data Availability}
The \textit{Planck} data used in this paper are available in the Planck Legacy Archive (\url{https://pla.esac.esa.int/}). The EDE data used in this paper were provided by Vivian Poulin by permission and, subject to his permission, will be shared on request to the corresponding author.

\section*{Acknowledgments} 
We thank Raphael Flauger, Katie Freese, Pawan Kumar, and Don Winget for helpful discussions and Vivian Poulin for providing EDE chains from \citet{murgia2021}. The \textit{Near/Far Napa Workshop (2019) helped to bring some of the ideas discussed here into focus for the authors.} MBK acknowledges support from NSF CAREER award AST-1752913, NSF grant AST-1910346, NASA grant NNX17AG29G, and HST-AR-15006, HST-AR-15809, HST-GO-15658, HST-GO-15901, HST-GO-15902, HST-AR-16159, and HST-GO-16226 from the Space Telescope Science Institute (STScI), which is operated by AURA, Inc., under NASA contract NAS5-26555. DRW acknowledges support from HST-AR-15006, HST-AR-15476, HST-GO-15901, HST-GO-15902, HST-AR-16159, and DD-ERS-1334 from STScI. DRW also acknowledges support from an Alfred P. Sloan Fellowship. 

The \textit{Planck} chains used in this paper are based on observations obtained with \textit{Planck} (\url{http://www.esa.int/Planck}), an ESA science mission with instruments and contributions directly funded by ESA Member States, NASA, and Canada. Much of the analysis in this paper relied on the python packages {\sc numpy} \citep{numpy2020}, {\sc scipy} \citep{scipy2020}, {\sc matplotlib} \citep{matplotlib}, {\sc ipython} \citep{ipython}, and {\sc GetDist} \citep{lewis2019}; we are very grateful to the developers of these tools. This research has made extensive use of NASA’s Astrophysics Data System (\url{http://adsabs.harvard.edu/}) and the arXiv e-Print service (\url{http://arxiv.org}).

\bibliography{draft.bbl}

\appendix
\section{Changes to cosmological quantities in EDE}
\label{sec:appendix}
%%%%%%%%%%%%%%%%%%%%%%%%%%%%%%%%%%%%%%%%%%%%%%%%%%%%%%%%%%%%%%
\begin{figure*}
 \centering
 \includegraphics[width=0.48\textwidth]{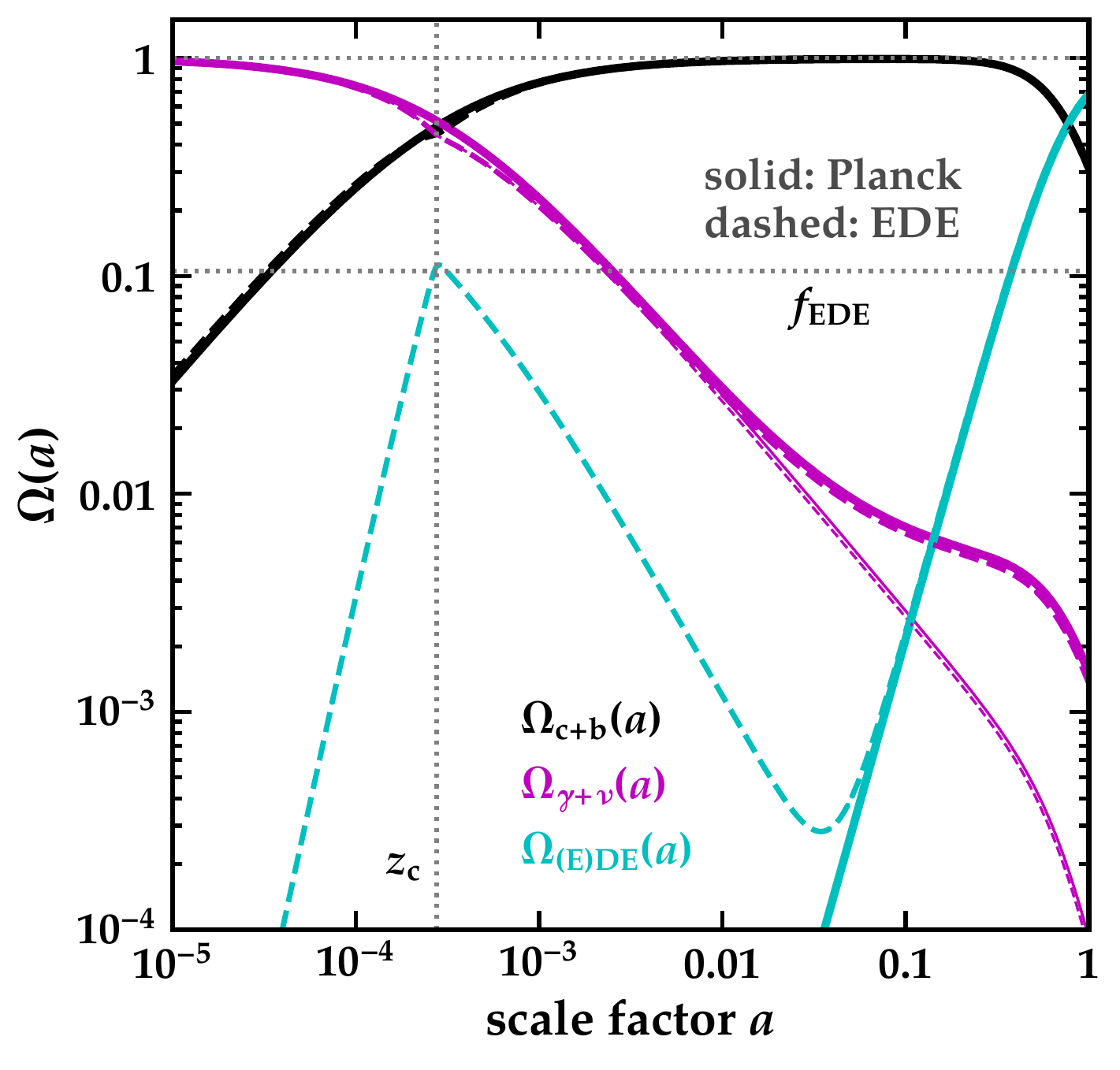}
  \includegraphics[width=0.48\textwidth]{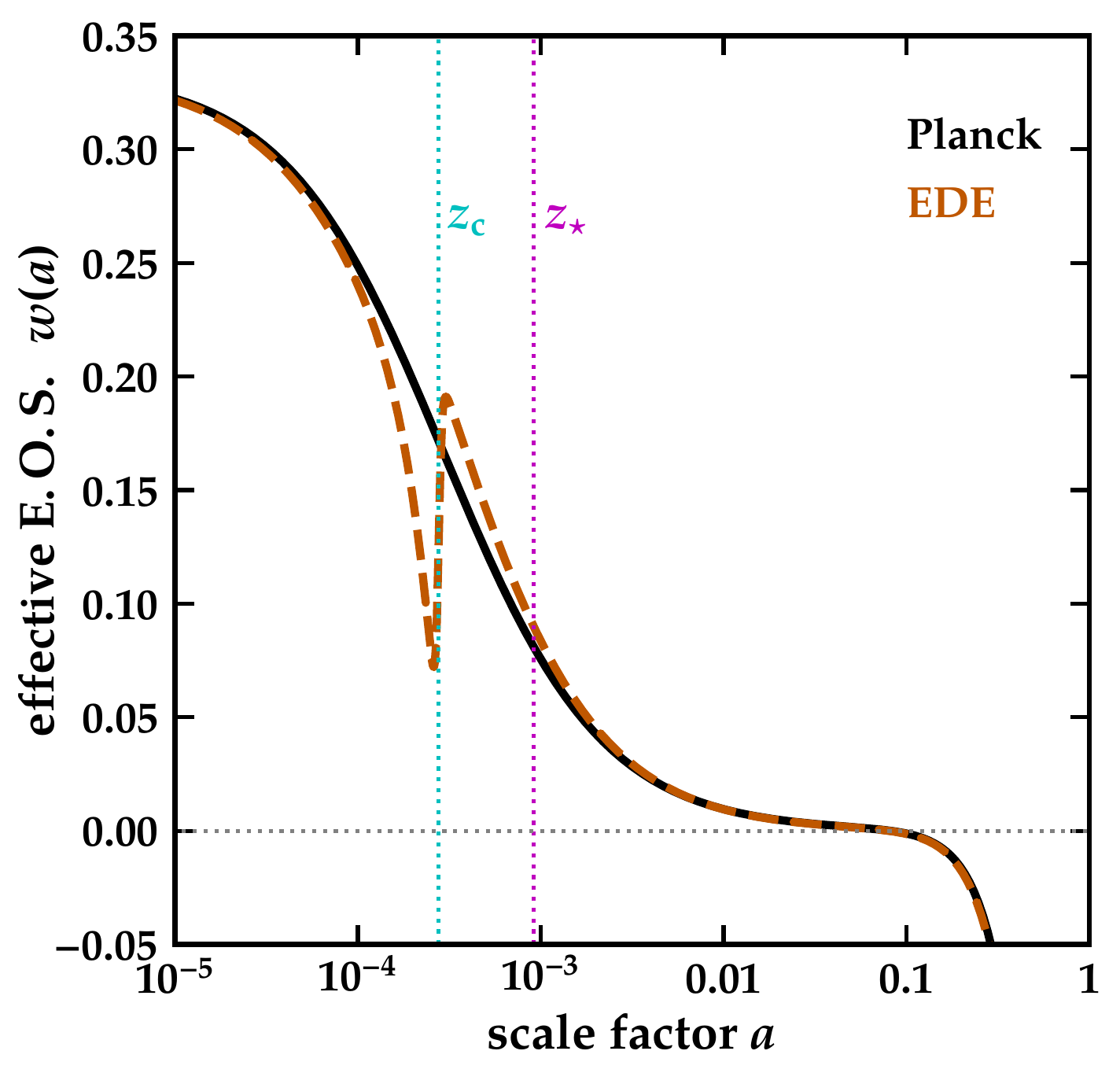}
 \caption{The effects of EDE on cosmological evolution of energy densities. \textit{Left:} density parameters for cold dark matter plus baryons (black), photons and neutrinos (magenta; the thin lines ignore neutrino mass), and dark energy (cyan) in \lcdm~(solid curves) and EDE (dashed curves). EDE peaks at a $\ssim 10\%$ contribution at $z_{\rm c}$ (see \citealt{klypin2021} for a version of this plot using the full Boltzmann code calculation for $\Omega_{\rm EDE}(a)$, which introduces small features at $a > a_{\rm c}$, rather than Eq.~\ref{eq:append_rho_ede}). \textit{Right:} The effective equation of state parameter $w_{\rm eff}(a)$, defined as via $d\ln H(a)/d\ln a=-3[1+w_{\rm eff}(a)]/2$, which governs the instantaneous change of the critical density $\rho_{\rm crit}(a)$. EDE follows the \planck~curve for $w_{\rm eff}(a)$ at early and late times, but near $z_{\rm c}$, it falls significantly below owing to the effects of the EDE component.
 \label{fig:appendix}
}
\end{figure*}
%%%%%%%%%%%%%%%%%%%%%%%%%%%%%%%%%%%%%%%%%%%%%%%%%%%%%%%%%%%%%% 
%%%%%%%%%%%%%%%%%%%%%%%%%%%%%%%%%%%%%%%%%%%%%%%%%%%%%%%%%%%%%%
\begin{figure*}
 \centering
 \includegraphics[width=0.48\textwidth]{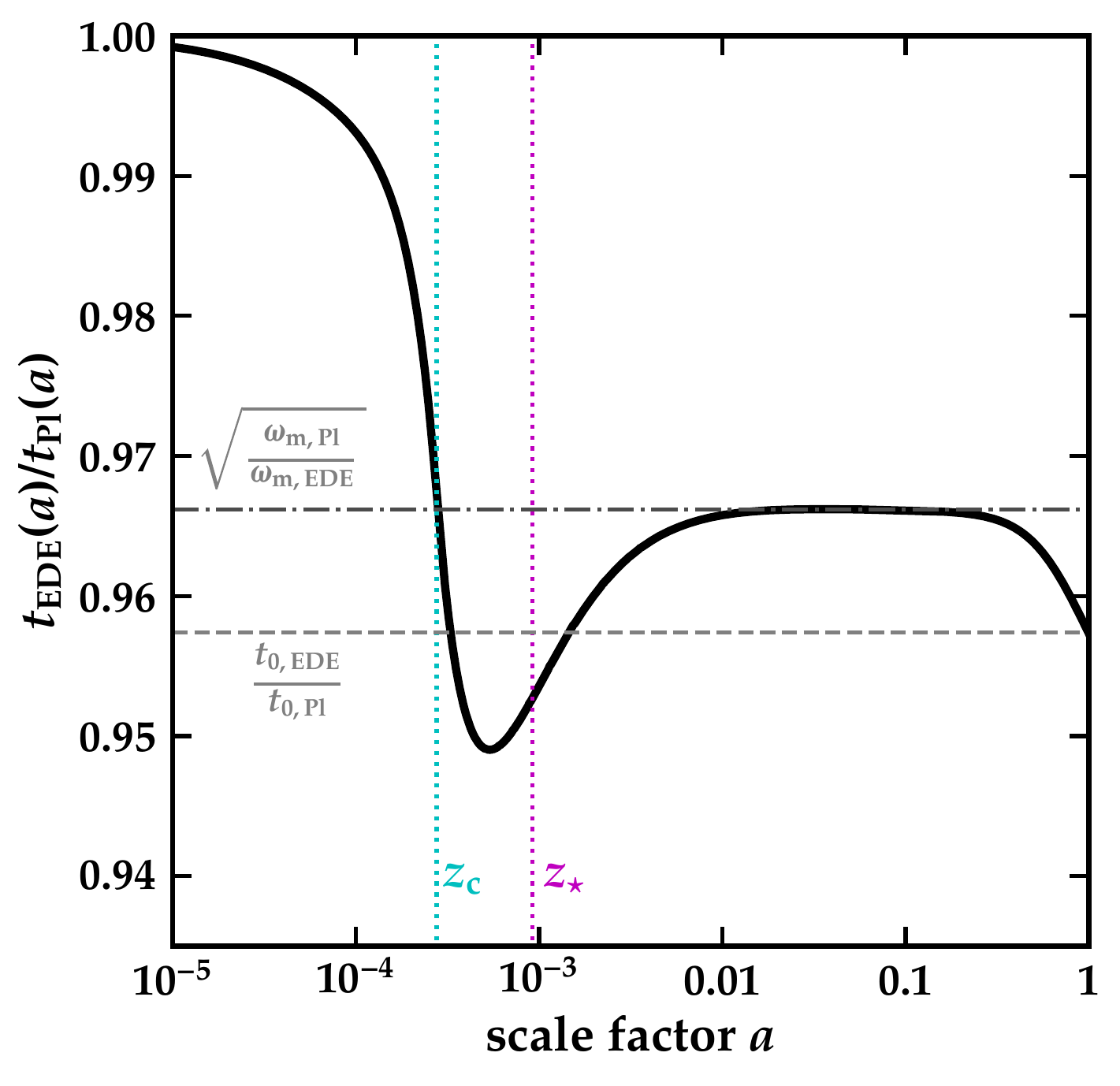}
 \includegraphics[width=0.48\textwidth]{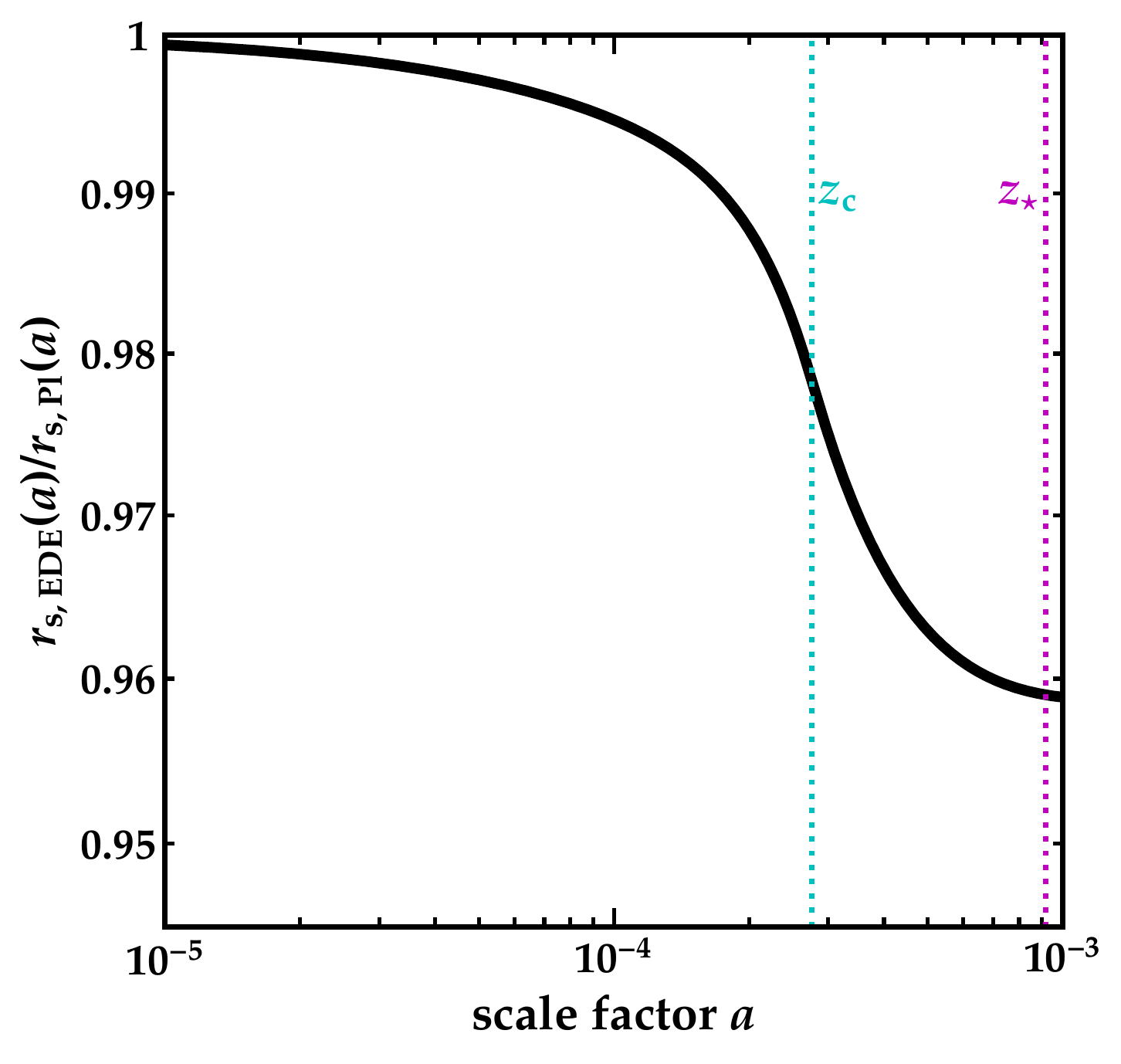}
 \includegraphics[width=0.50\textwidth]{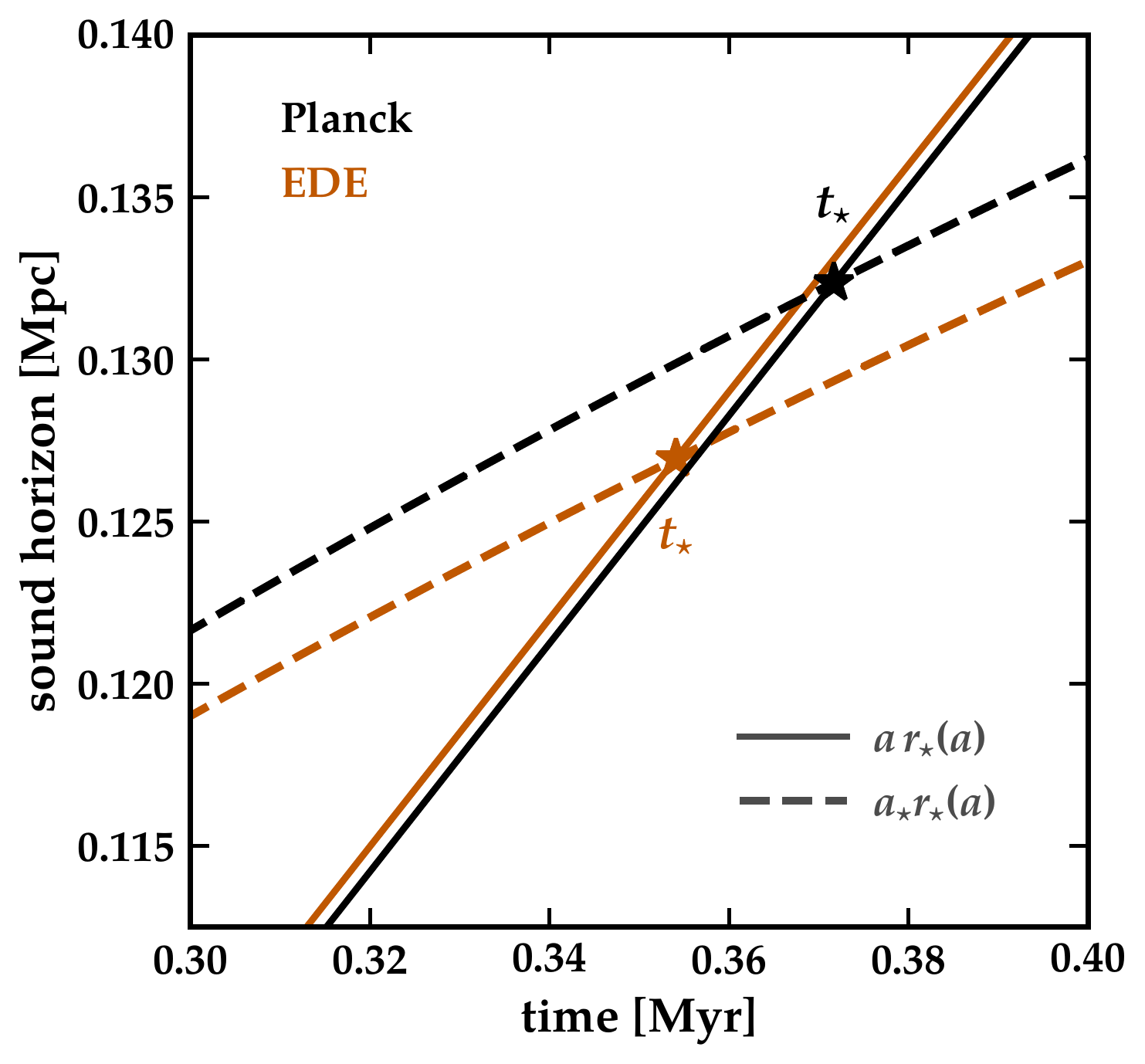}
  \caption{The relationship between time and scale factor (upper left panel), sound horizon and scale factor (upper right panel), and sound horizon and time (lower panel) for \planck~and EDE. The small component of EDE leads the Universe to reach a fixed scale factor at an earlier time. Combined with the very nearly identical sound speed in the photon-baryon plasma for the two models, this means that the sound horizon at a given scale factor is smaller in EDE than in \planck. The sound horizon as a function of time (lower panel) shows that the physical sound horizon (solid curves) is slightly larger at fixed $t$ in EDE, as expected from the existence of an extra dark energy component. Note, however, that the comoving sound horizon --- shown in dashed curves, scaled by $a_{\star}$ --- is smaller in EDE at a fixed time. The change in the expansion history for EDE means that $a_{\star}$ occurs earlier, making the sound horizon 4\% smaller in EDE than in \planck.
 \label{fig:appendix_rsound}
}
\end{figure*}
%%%%%%%%%%%%%%%%%%%%%%%%%%%%%%%%%%%%%%%%%%%%%%%%%%%%%%%%%%%%%% 

While the results of this paper are not focused on a specific EDE model (or even EDE itself) but rather the effects of a modified expansion history on the redshift-time relation, it is nonetheless useful to understand how a short period of early accelerated expansion changes the sound horizon at last scattering and related quantities. 

The results of \S\S~\ref{sec:cosmology} and \ref{sec:cosmo_ages} are mostly derived under the assumption of the base \lcdm~model. In a model with additional energy content in some component X, the calculations of ages and distances need to be modified because the expansion history is modified. For models that modify only the early-time ($z > z_{\star}$) expansion history, only ages and times in the early Universe ($z \gtrsim 0.1\,z_{\star}$) are affected. This change simply requires replacing Eq.~\ref{eq:hub} with a modified version, 
\begin{equation}
\label{eq:appendix_hz}
    H_{\rm }(a)=H_0\,\sqrt{\om\,a^{-3} + (1-\om)+\orad\,a^{-4}+\Omega_{\rm X}\,g(a)} \,,
\end{equation}
which takes into account the present-day density in component X (encapsulated in the density today relative to the critical density today, $\Omega_{\rm ede}$) and its evolution with scale factor, $g(a)=\rho_{\rm X}(a)/\rho_{\rm X}(a=1)$. 

Eq.~\ref{eq:appendix_hz} is general; different types of energy will have different versions of $g(a)$, often with $g(a) \propto a^{-3(1+w_{\rm X})}$ for equation of state (EOS) parameter $w_{\rm X}(a)=P_{\rm X}(a)/\rho_{\rm X}(a)$ encapsulating the ratio of the pressure to the density in component X at scale factor $a$.
The density evolution with time for the EDE model considered here can be approximated by 
\begin{equation}
\label{eq:append_rho_ede}
    g(a)=\bigg[\frac{1+(1/a_{\rm c})^{3(1+w_{\rm n})\xi}}{1+(a/a_{\rm c})^{3(1+w_{\rm n})\xi}}\bigg]^{1/\xi}\,,
\end{equation}
with $a_{\rm c}=(1+z_{\rm c})^{-1}$ and $w_{\rm n}=(n-1)/(n+1)$; for the $n=3$ model considered here, $w_n=\nicefrac{1}{2}$. Eq.~\ref{eq:append_rho_ede} is identical to the fit to density evolution given in \citetalias{poulin2018} (their Eq.~15]) for $\xi=1$. Larger values of the empirical factor $\xi$ correspond to a faster transition between the EDE phase and the decay of EDE (see fig.~2 of \citetalias{poulin2018}); we adopt $\xi=10$, which provides a closer approximation to the full numerical calculation for $n=3$ from \citetalias{poulin2018} than their eq.~15. For $a \gg a_{\rm c}$, $\rho \propto a^{-3(1+w_{\rm n})}=a^{-9/2}$ for $n=3$, while for $a \ll a_{\rm c}$, $g(a) \approx {\rm constant}$. The value of $\Omega_{\rm EDE}(a=1)$ can be computed in terms of $\Omega_{\rm EDE}(a_{\rm c})=f_{\rm EDE}$ (one of the primary parameters of the EDE model) from Eq.~\ref{eq:append_rho_ede} and is equal to
\begin{equation}
    \label{eq:appendix_omega_ede}
    \Omega_{\rm EDE}=\frac{f_{\rm EDE}}{1-f_{\rm EDE}}\,\frac{1}{g(a_{\rm c})}\,\left(\Omega_{\rm m}\,a_{\rm c}^{-3}+\Omega_{\rm r}\,a_{\rm c}^{-4}+\Omega_{\Lambda} \right)\,.
\end{equation}

Figure~\ref{fig:appendix} shows the evolution of $\Omega_{\rm i}(a)=\rho_{\rm i}(a)/\rho_{\rm crit}(a)$ [left] and the effective EOS $w_{\rm eff}(a)$ [right], defined via
\begin{equation}
    \label{eq:appendix_weff}
    H(a)=H_0\,a^{-3(1+w_{\rm eff})/2}\,,
\end{equation}
for \planck~(solid curves) and EDE (dashed curves). The left panel shows the usual periods of dark energy, matter, and radiation domination remain mostly unchanged when moving from \planck~to EDE. The notable difference is that at $z_{\rm c}$, EDE contributes approximately 10\% of the critical energy density; at lower and higher redshift, it contributes less. This small contribution from EDE --- which coincides with $z_{\rm eq}$ --- pushes $z_{\rm eq}$ to a slightly higher redshift and changes the expansion rate as a function of scale factor. Notably, the Hubble expansion rate as a function of \textit{time} in EDE remains nearly identical to the base \lcdm~model. The effective EOS (right panel of Fig.~\ref{fig:appendix}), which encapsulates how $\rho_{\rm crit}(a)$ is changing with scale factor, is identical in the two models for $a \ll a_{\rm c}$ and $a \gg a_{\rm c}$. Near $z_{\rm c}$, however, the EDE $w_{\rm eff}$ drops sharply, reflecting the increase in importance of EDE, followed by an abrupt jump back to a larger value owing to the decay of EDE. Since EDE dilutes as $\rho \propto a^{-9/2}$ --- $w_{\rm EDE} = \nicefrac{1}{2}$ --- for $a > a_{\rm c}$, the effective EOS in EDE is slightly larger than for \planck~until the EDE energy density has become completely negligible compared to radiation.

Figure~\ref{fig:appendix_rsound} demonstrates how EDE modifies the cosmological  relationship between distance, time, and scale factor in the early Universe. The effect of EDE on the relationship between scale factor and time is that the EDE universe reaches a fixed scale factor at an earlier time because of the component of accelerated expansion --- note that the ratio $t_{\rm EDE}/t_{\rm Pl}$ starts at unity in the period before EDE becomes dynamically significant. At $a_{\star}$ --- which, crucially, does not change between the two models for the relatively low values of $f_{\rm EDE}$ being considered --- the EDE universe is 4.7\% younger than the \planck~universe. A sound wave with a given velocity $c_{\rm s}(t)$ will therefore not travel as far in EDE as in \planck, reducing $r_{\star}$ in EDE by 4.1\% relative to \planck. (The sound speed differs very slightly between the two models, but this is only a 0.3\% effect at $a_{\star}$; this difference declines at smaller scale factors and is therefore always subdominant to the change in $a(t)$ between the two models.) While the \textit{physical} sound horizon is slightly larger at a given time in EDE than in \planck\ --- as one might expect for a Universe that is has a component that is acting as dark energy --- this is overwhelmed by the change in $a(t)$, meaning the comoving sound horizon is larger in \planck~at fixed time. The change in $t(a_{\star})$ between the two models further increases the difference in the comoving sound horizon.

\end{document}

%% file: extra_preamble.tex
\newcommand{\msun}{M_{\odot}}

\newcommand{\mpc}{{\rm Mpc}}

\newcommand{\kms}{{\rm km \, s}^{-1}}
\newcommand{\gyr}{{\rm Gyr}}

\newcommand{\lcdm}{$\Lambda$CDM}
\newcommand{\bmlcdm}{$\mathbf{\Lambda}$CDM}